\documentclass{aa}%

\usepackage{graphicx}
\usepackage{txfonts}
\usepackage{hyperref}
\begin{document}

   \title{The effect of winds on atmospheric layers of red supergiants}

   \subtitle{II. Modelling VLTI/GRAVITY and MATISSE observations of AH~Sco, KW~Sgr, V602~Car, CK~Car, and V460~Car   
\thanks{Based on observations made with the Very Large Telescope  Interferometer at Paranal Observatory under programme IDs 101.D-0616, 109.231U, and 110.23P1}}

\author{G. Gonz\'alez-Tor\`a
          \inst{1,2,3}
          \and
          M. Wittkowski\inst{2}
          \and
          B. Davies\inst{3}
          \and
          B. Plez\inst{4}
        }
   
   \institute{Zentrum für Astronomie der Universität Heidelberg, Astronomisches Rechen-Institut, 
   Mönchhofstr. 12-14, 69120 Heidelberg\\
        \email{gemma.gonzalez-tora@uni-heidelberg.de}
   \and
   European Southern Observatory (ESO),
   Karl-Schwarzschild-Stra{\ss}e 2, 85748 Garching bei München, Germany\
   \and
        Astrophysics Research Institute, Liverpool John Moores
        University, 146 Brownlow Hill, Liverpool L3 5RF, United Kingdom\
    \and
       LUPM, Universit\'e de Montpellier, CNRS, Montpellier, France}  

   \date{Received \today; accepted -}

% \abstract{}{}{}{}{} 
% 5 {} token are mandatory
 
  \abstract
  %\LEt{ General notes: A.) I have edited to UK English spelling and grammar conventions. B.) A\&A uses the past tense to describe specific methods used in a paper and the present tense to describe general methods as well as findings, including the findings of recent papers (within the past ten or so years). See Sect. 6 of the language guide https://www.aanda.org/for-authors/language-editing/6-verb-tenses. ***} 
   { Mass loss plays a crucial role in the lives of massive stars, especially as the star leaves the main sequence and evolves to the red supergiant (RSG) phase.  Despite its importance, the physical processes that trigger mass-loss events in RSGs are still not well understood. Recently, we showed that adding a semi-empirical wind to atmosphere models can accurately reproduce observed extensions in the atmospheres of RSGs, where the mass-loss events are taking place, particularly in the CO and water lines. %Recently -> in Chapter~\ref{ch:amber}
   }
   { By adding a static wind to a MARCS atmospheric model, we computed synthetic observables that match new interferometric data of the RSGs AH~Sco, KW~Sgr, V602~Car, CK~Car, and V460~Car obtained with the VLTI/MATISSE and VLTI/GRAVITY instruments between August 2022 and February 2023. We also used archival VLTI/AMBER data of KW~Sgr and VLTI/GRAVITY data of AH Sco. The MATISSE wavelength range includes the SiO molecule at $4.0\,\mu$m with a spectral 
   resolution of $R\sim500$. 
   }
{The model intensities with respect to the line-of-sight angle ($\mu$) as well as the spectra and visibilities were computed using the stellar radiative transfer code \textsc{Turbospectrum}. We found the best-fit model, mass-loss rate, and best-fit angular Rosseland diameter for the observations. We simultaneously matched our model to the data, covering a wavelength range of $1.8-5.0\,\mu$m, which corresponds to the $K$, $L,$ and $M$ bands.}
   { Our models reproduce the spectro-interferometric data over this wide wavelength range, including extended atmospheric layers of CO, H$_2$O, and SiO. 
   We obtain a range of Rosseland angular diameters between $3.0<\theta_{\mathrm{Ross}}<5.5$ mas and a range of mass-loss rates of $-6.5<\log \dot{M}/M_{\odot}\mathrm{yr}^{-1}<-4$ for our five targets. In our best-fit models, the partial pressure of SiO relative to the gas pressure, $P_\mathrm{SiO}/P_\mathrm{g}$, 
   and the SiO 4.0\,$\mu$m line intensity increase between 2 and 3 stellar radii. The relative intensity depends on the luminosity used
for our models, since the more luminous models have a higher mass-loss rate.
   }
   { This work further demonstrates that our MARCS+wind model can reproduce the observed physical extension of RSG atmospheres for several spectral diagnostics spanning a broad wavelength range. We reproduce both spectra and visibilities of newly obtained data as well as provide temperature and density stratifications that are consistent with the observations. With the MATISSE data, we newly include the extension of SiO layers as a precursor of silicate dust.% Moreover, this work also represents one of the first most complete studies of the atmospheric structure of RSGs done with GRAVITY and MATISSE data to date. %A simple temperature profile seems to reproduce well the full wavelength coverage explored, going against the initial lukewarm chromospheres explanation.  
  }

   
   \maketitle
%
%-------------------------------------------------------------------

\section{Introduction}\label{sec:intro}

Red supergiants (RSGs) are evolved massive stars in the stage before core-collapse supernovae. These stars extend their atmosphere by several stellar radii. Farther out, the star is surrounded by a dust shell. RSGs incur mass-loss events in their extended atmospheres. Arguably, one of the most famous mass-loss episodes to date was the great dimming of Betelgeuse in early 2020 \citep{2019ATel13341....1G,2020ApJ...899...68D, 2020ApJ...897L...9D}, when its visual brightness decreased by $0.6$ magnitudes. Various studies revealed that the most likely cause of its dimming was a mass-loss event in the southern hemisphere of the star \citep{2021Natur.594..365M,2022ApJ...936...18D}. However, the processes that trigger these mass-loss events are still not well understood. 

The Betelgeuse mass-loss event may have been related to a stochastic occurrence of an extreme convection cell enhanced by pulsation 
\citep{2020ApJ...899...68D,2020ApJ...905...34H,2022ApJ...936...18D}, a process that may also explain the mass-loss history of VY CMa \citep[][]{2021AJ....161...98H} and of RSGs in general \citep[][]{2022AJ....163..103H}. These RSGs appear to have enhanced directed outflows combined with a steady wind. 
As a consequence of the poorly understood physical mechanisms, few studies have modelled this mass-loss effect in cool massive stars: most have focused only on mid- and far-IR regions, where the dust component plays a major role \citep[e.g.][]{2018MNRAS.475...55B}. 

Current modelling attempts to reproduce atmospheric extension include the addition of a thin static molecular shell \citep[or 'MOLsphere'; e.g.][]{2007A&A...474..599P,2009A&A...503..183O,2011A&A...529A.163O,2014A&A...572A..17M,2018A&A...609A..67K} and 3D radiative hydrodynamics studies that are not sufficiently extended for RSGs \citep[e.g.][]{2009A&A...506.1351C,2010A&A...515A..12C,2011A&A...528A.120C,2011A&A...535A..22C}. Recent work by \mbox{\citet{2021A&A...646A.180K}} studied the effect of turbulent atmospheric pressure in initiating and determining the mass-loss rates ($\dot{M}$) of RSGs, finding promising results in retrieving the observed $\dot{M}$. However, there is still no dynamical model available that succeeds in reproducing mass-loss events \citep{2015A&A...575A..50A}. Therefore, further work is needed to unambiguously determine the dynamical processes that trigger massive star wind events in spatially extended atmospheres. 
%\textit{can I use the same sentence as the other paper?}

Recently, \citet{2021MNRAS.508.5757D} developed a semi-empirical model based on a hydrostatic MARCS model atmosphere \citep{2008A&A...486..951G} with the addition of an empirical wind profile. Using this model to extend the stellar atmosphere, \citet{2021MNRAS.508.5757D} show that the variability of Betelgeuse during the great dimming did not require dust: while the optical flux dipped by an order of magnitude, the near-IR flux remained constant. This phenomenon can be explained by the optical depth increase in the TiO bands using the model from \citet{2021MNRAS.508.5757D}. %reproduce most of the spectral features 
%as obtained by adding a separate MOLsphere, 
%without the need of dust

In the first paper of this series \citep[][hereafter Paper I]{2023A&A...669A..76G}, we compared the model approach taken by \citet{2021MNRAS.508.5757D} to published Very Large Telescope Interferometer (VLTI) Astronomical Multi-BEam combineR (AMBER) data from \citet{2015A&A...575A..50A}. Thanks to its unprecedented high angular spatial resolution, interferometry allows us to explore the extension of the atmospheres of RSGs by resolving the stellar disk. Paper I shows that our models reproduce the atmospheric extension for the first time up to $\sim 8R_{\star}$, where $R_{\star}$ is the stellar radius at the photosphere, defined as the layer where the Rosseland optical depth is $\tau_{\mathrm{Ross}}=2/3$. The squared visibility amplitudes ($|V|^{2}$), observables of the interferometric data, could be reproduced as well for the first time in the $K$ band. These important results open a window to exploring mass-loss events of cool massive stars since they provide atmospheric extensions and constrain the temperature and density stratification.% and there is no need to rely anymore only on the external dust component.
    
In addition, Paper I introduced the concept of lukewarm chromospheres: a heterogeneous environment with cells of different temperatures at the same radii. It is likely that the RSGs studied in this paper have similar heterogeneous structures. %the episodic outflows is mentioned in the 2nd paragraph

For this work, we obtained new VLTI Multi AperTure mid-Infrared SpectroScopic Experiment (MATISSE) data of the RSGs AH~Sco, V602~Car, CK~Car, V460~Car, and KW~Sgr in the thermal IR, which allows us to include in our
study the extension of the SiO line at 4.0\,$\mu$m as well as CO lines in the $M$ band. We also obtained GRAVITY data for the aforementioned targets 
in the $K$ band, except for KW~Sgr, whose data we complemented with VLTI/AMBER data published by \citet{2013A&A...554A..76A}. We simultaneously analysed both datasets and obtained an estimation of the best-fit $\dot{M}$ from our static stellar wind model, as well as temperature, density, and pressure stratification profiles and Rosseland angular diameters. 
 
\medskip
This paper is organised as follows: In Sect.~\ref{sec:obs} we briefly present the instruments used to make the observations, as well as the data reduction process. The model and analysis are introduced in Sect.~\ref{sec:methods}. The results for our observations are presented in Sect.~\ref{sec:results}. We discuss the implications of our results in Sect.~\ref{sec:diss}. Finally, we conclude in Sect.~\ref{sec:conc}.

%--------------------------------------------------------------------
\section{Observations and data reduction}\label{sec:obs}

To obtain spectro-interferometric data, we used two instruments of  the VLTI: MATISSE and GRAVITY. In the following, we briefly introduce the main characteristics of both instruments.

\subsection{VLTI observations}

GRAVITY is an interferometer of the VLTI \citep{2017A&A...602A..94G} optimised for the $K$ band, comprising the wavelength range of $1.8<\lambda<2.4\,\mu$m. 
The GRAVITY instrument has two different modes: the split polarisation mode splits the light into two polarisation angles and increases the internal fringe contrast of the instrument, the combined polarisation mode increases the sensitivity instead. Our targets are bright enough, so we chose the split polarisation mode. 

%MATISSE
MATISSE \citep[][]{2022A&A...659A.192L} is a mid-IR spectro-interferometer that combines four beams of the VLTI. The instrument is optimised for the $L$ ($2.8<\lambda<4.1\,\mu $m), $M$ ($4.5<\lambda<5\,\mu $m) and $N$ ($8<\lambda<13\,\mu $m) bands.
MATISSE uses a beam combiner, where the four separated telescope beams are focused in a detector, producing six dispersed fringe patterns mixed into a single focal spot. For the $L$ and $M$ bands, MATISSE uses the SiPhot photometric mode, where the photometry is measured simultaneously with the dispersed interference fringes. This mode allows the proper calibration of the visibilities. For the $N$ band, the photometric measurement mode is obtained separately after the interferometric observations, allowing the observer to improve the sensitivity by sending the total flux in the different channels (HiSens or High Sensitivity mode). Recently, a new mode designated for faint targets uses the GRAVITY instrument as a fringe-tracker, called GRAVITY for MATISSE \citep[or GRA4MAT;][]{2022A&A...659A.192L}. This new mode provides improved sensitivities for the auxiliary telescopes (ATs) and increases the spectral coverage.

We observed the RSGs KW~Sgr (ID: 109.231U.001, PI: Wittkowski), AH-Sco (ID: 109.231U.002, PI: Wittkowski), V602~Car, CK~Car, and V460~Car (ID: 110.23P1.001, PI: Gonz\'alez-Tor\`a) with VLTI/MATISSE. We also observed AH~Sco (ID: 0101.D-0616(B), PI: Wittkowski), V602~Car, CK~Car, and V460~Car (ID: 110.23.P1.002, PI: Gonz\'alez-Tor\`a) with VLTI/GRAVITY.

For the GRAVITY observations ($1.8<\lambda<2.4\,\mu$m), the targets were observed as snapshots with single split-polarisation mode and the medium interferometric baseline configuration. We used the HIGH spectral resolution $R\sim4000$ mode. The sequences were observed as CAL-SCI-CAL, meaning we first observed a calibrator, then the science target and again a calibrator. For each of these three observations, the sequence was OOSOOS, where O denotes the object (SCI or CAL) and S the sky position. 

For the MATISSE observations ($2.8<\lambda<13\,\mu$m), the targets were observed as snapshots with the medium interferometric baseline configuration. For the $L$ and $M$ bands we used the MED spectral resolution $R\sim500$, while for the $N$ band the HIGH $R\sim1000$ for AH~Sco and KW~Sgr, and LOW $R\sim30$ for V602~Car, CK~Car, and V460~Car. We observed the sequences CAL-L, SCI, CAL-N for the same night, where CAL-L is the calibrator for the $L$ and $M$ bands, SCI is the science target and CAL-N is the calibrator for the $N$ band. Usually, we cannot use the same calibrators for the $L/M$ and $N$ bands, since we need specific angular diameters and magnitudes that depend on the bands we are observing. Our observations were all taken with the new GRA4MAT mode, to obtain the full wavelength coverage of $2.8<\lambda<5\,\mu$m at medium spectral resolution in the $L$ and $M$ bands.

The date and time of the observations, as well as the observing conditions are shown in Tables~\ref{tab:obsg} and \ref{tab:obsm} for GRAVITY and MATISSE, respectively. The stellar properties of our observed calibrators are shown in Table~\ref{tab:calib}. 

The targets studied are variable stars, whose variability between times of observations could affect the results. We compared the observational dates with the enhanced light curves available from the {American Association of Variable Star Observers}\footnote{\url{https://www.aavso.org/LCGv2/}}. For the case of AH~Sco and KW~Sgr, the observations were taken around the minimum of the variability cycles. For the other RSGs, the observations were taken within a month, so that the variability of the stars in this short period would not affect our results. 

\begin{table*}
\caption{Summary of the VLTI/GRAVITY data obtained for the five observed RSGs and their calibrators. Shown are the target name, the date when the observation was taken, the UT time at the start of the observation, the seeing of the science target, the coherence time, $\tau_{0}$, at the start of the observation, the calibrator name, the seeing of the calibrator, the coherence time of the calibrator, and the AT configuration.  }
\label{tab:obsg}
\small
\centering
\begin{tabular}{c c c c c c c c c c}
    \hline \hline
      Target & Date & UT & Seeing sci ($\arcsec$) & $\tau_{0}$ (ms) &
      Calibrator & Seeing calib ($\arcsec$) & $\tau_{0}$ (ms) 
      & AT configuration \\
     \hline %\hline
      AH Sco & 17-06-18 & 01:21:25 & 1.09 & 2.23 & HD159881 & 1.05 & 3.60 & K0G2D0J3/Medium \\
       & & & & & HD152636 & 0.89 & 2.23 & K0G2D0J3/Medium \\
       %     \hline
      V602 Car & 22-01-23 & 05:23:32 & 0.65 & 8.20 & HD96442 & 0.7 & 7.94 & K0G2D0J3/Medium \\
       & & & & & HD103859& 0.63 & 7.84 & K0G2D0J3/Medium \\
        %    \hline
     CK Car & 22-01-23 & 04:34:00 & 0.68 & 7.5 & HD89736 & 0.71 & 6.9 & K0G2D0J3/Medium \\
       & & & & & HD90677 & 0.82 & 6.1 & K0G2D0J3/Medium \\
   % \hline
          V460 Car & 25-11-22 & 05:40:34 & 0.75 & 9.1 & Q Car & 1.06 & 6.5 & K0G2D0J3/Medium \\
       & & & & & HD 60228 & 0.65 & 11.1 & K0G2D0J3/Medium \\
     \hline
     %     VAR Cet & 23-11-22 & 01:31:10 & 0.74 & 6.94 & ups Cet & 0.56 & 7.78 & K0G2D0J3/Medium \\
      % & & & & & 80 Cet & 0.6 & 9.14 & K0G2D0J3/Medium \\
     %\hline
\end{tabular}

\end{table*}

\begin{table*}
\caption{Summary of the VLTI/MATISSE data obtained for the five observed RSGs and their calibrators. The columns give the target name, the date when the observation was taken, the UT time at the start of the observation, the seeing of the science target, the coherence time, $\tau_{0}$, at the start of the observation, the calibrator name, the seeing of the calibrator, the coherence time of the calibrator, and the AT configuration.  }
\label{tab:obsm}
\small
\centering
\begin{tabular}{c c c c c c c c c c}
        \hline \hline
      Target & Date & UT & Seeing sci ($\arcsec$) & $\tau_{0}$ (ms) &
      Calibrator & Band & Seeing calib ($\arcsec$) & $\tau_{0}$ (ms) 
      & AT configuration \\
     \hline %\hline
        AH Sco & 26-08-22 & 23:45:59 & 0.43 & 7.2 & gam02 Nor & L\&M & 0.81 & 6.7 & K0G2D0J3/Medium  \\
       & & & & & eps Sco & N & 0.33 & 8.0 & K0G2D0J3/Medium  \\
     %  \hline
      KW Sgr & 28-08-22 & 00:16:46 & 0.43 & 3.1  & nu Oph & L\&M & 0.65 & 3.5 & K0G2D0J3/Medium \\
       & & & & & eps Sco & N & 0.45 & 4.2 & K0G2D0J3/Medium \\
    % \hline
    V602 Car & 19-01-23 & 06:57:36 & 0.65 & 1.62 & HD94683 & L\&M & 0.73 & 1.97 & K0G2D0J3/Medium \\
       & & & & & HD91056 & N & 0.51 & 1.91 & K0G2D0J3/Medium \\ 
    % \hline
    CK Car  & 27-02-23 & 03:16:25 & 0.46 & 9.4 & HD94683 & L\&M & 0.88 & 8.3 & K0G2D0J3/Medium \\
       & & & & & HD302821 & N & 0.44 & 13.2 & K0G2D0J3/Medium \\
     %  \hline
        V460 Car & 26-12-22 & 04:06:43 & 1.16 & 2.5 & zet Vol & L\&M & 0.93 & 2.9 & K0G2D0J3/Medium \\
       & & & & & CD-55 3254 & N & 0.86 & 2.9 &K0G2D0J3/Medium \\
       \hline
       %    VAR Cet & 25-11-22 & 03:07:10 & 0.61 & 1.24 & HD16074 & L\&M & 0.89 & 6.38 & K0G2D0J3/Medium \\
      % & & & & & ups Cet & N & 0.67 & 1.52 & K0G2D0J3/Medium \\
      % \hline
\end{tabular}

\end{table*}

\begin{table*}
\caption{Properties of the calibrators for the VLTI/MATISSE and GRAVITY observations. Shown are the name, spectral type, limb-darkened disk diameter (LDD) from \citet{2017yCat.2346....0B}, and the stellar parameters of the synthetic spectra \citep[MARCS,][]{2008A&A...486..951G} used for the calibrators: effective temperature ($T_{\mathrm{eff}}$), surface gravity ($\log g$), and metallicity [Z] available in the VizieR database of astronomical catalogues \citep{2000A&AS..143...23O}. }
\label{tab:calib}
\small
\centering
\begin{tabular}{c c c c c c }
\hline \hline
      Name & Spectral type & LDD diameter (mas) & $T_{\mathrm{eff}}$ (K) & $\log g$ & [Z]  \\
     \hline %\hline
        gam02 Nor & K0III & 2.46$\pm$0.25 & 4750 & 3.0 & 0.25 \\
 %    \hline
        eps Sco & K1III & 5.93$\pm$0.45 & 4500 & 2.5 & -0.25  \\
 %            \hline
        nu Oph & K1III & 2.83$\pm$0.29 & 5000 & 3.0 & 0.00  \\
%             \hline
        HD94683 & K4III & 2.49$\pm$0.25 & 4000 & 0.0 & 0.00 \\
   %          \hline
        HD91056 & M0III &  4.40$\pm$0.49 & 3800 & 1.0 & 0.00 \\
   %          \hline
       HD16074 & K2III &  1.77$\pm$0.13 & 4250 & -0.5 & 0.00 \\
  %          \hline
      ups Cet & M0III &  5.91$\pm$0.76 & 3800 & 1.15 & -0.05 \\
      
      HD302821 & A7Ie & 1.26$\pm$0.10  & 3700 & 0.00 & 0.00 \\
            
      zet Vol & K0III & 2.23$\pm$0.21 & 4700 & 2.5 & -0.2 \\
                  
      CD-55 3254 & M3 & 2.76$\pm$0.27 & 3300 & 0.00 & 0.00 \\
     %\hline 
    % \hline
       HD159881 & K5III &  2.81$\pm$0.24 & 4500 & 2.5 & -0.25 \\
 %          \hline
       HD152636 & K5III &  2.39$\pm$0.22 & 4000 & 2.5 & -0.05 \\
  %         \hline
      HD96442 & M1/2III &  1.64$\pm$0.16 & 4000 & 0.00 & 0.00 \\
 %          \hline
      HD103859 & K4III &  1.43$\pm$0.11 & 4000 & 0.00 & 0.00 \\
 %          \hline
      80 Cet & M0III &  3.57$\pm$0.36 & 4000 & 0.00 & 0.00 \\
      
      HD89736 & K5/M0III & 2.58$\pm$0.23 & 3900 & 0.00 & 0.00 \\
      
      HD90677 & K3II & 1.99$\pm$0.19 & 4000 & 0.00 & 0.00 \\
      
      Q Car & K3III &  2.45$\pm$0.27 & 4300 & 0.00 & 0.00 \\
      
      HD60228 & M1III & 2.51$\pm$0.24 & 3900 & 0.00 & 0.00 \\
 %          \hline
%      ups Cet &  &  $\pm$ & 3800 & 1.15 & -0.05 \\
     \hline
 %parameters from wikipedia, LDD and SPTfrom Bourges+17    
\end{tabular}

\end{table*}

\subsection{Data reduction}
For the data reduction, we used the \textsc{ESOreflex} workflows with the GRAVITY pipeline version 1.6.0 and the MATISSE pipeline version 1.7.6\footnote{The reduction pipelines are available at \url{http://www.eso.org/sci/software/pipelines/}}. The steps of the data reduction for both instruments are described in \citet{2014SPIE.9146E..2DL} and \citet{2016SPIE.9907E..23M}, respectively. %Briefly: the master calibration files are created (e.g., flat field, kappa matrix) and the science reduction stars by compensating some offsets and dividing by the flat field. %after the data has been organized and selected, the master calibration files are created: the flat field is obtained, a distortion estimation (or shift map) is combined, the kappa matrix is estimated. The kappa matrix is the linear transformation matrix of the intensities in the photometric channels into the interferometric channel.
%Then, the science reduction stars: compensate for some offsets (e.g., pixel bias), compensate for non-linearity intensity response, divide by the flat field map, interpolate bad detector pixels in each frame, apply a distortion matrix to transform the coordinates, adjust photometric contributions by applying the kappa matrix, shift and zoom coefficients and remove the thermal background with chopping the photometric beams estimate. 
%Afterwards, the fluxes are calculated by preforming a Fourier transform \citep{2003A&A...400.1173P}. 
%Subsequently, the optical path difference (OPD) is calculated. The OPD modulation is aimed at removing the low-frequency peak contamination, containing the contribution from the thermal background to the fringe peaks. 
The $|V|^{2}$, the visibilities, closure phases, differential phases and coherent fluxes are calculated by the pipeline. 

Once the raw data has been reduced, the second \textsc{ESOreflex} workflow calibrates the visibilities of the science data using the different calibrators. For MATISSE, the calibration workflow changes slightly for both $L/M$ and $N$ bands: for the $N$ band, we changed the spectral binning to 11 bins/$\lambda$ and the coherent integration time to 0.3 s. %The initial values are 7 bins/$\lambda$ for spectral binning and 0.0 s for the coherent integration time, used in the $K$ band. As mentioned, the calibrators are different depending on the band we are reducing.

{\sc ESOreflex} does not provide us with a flux calibration method; therefore, we calibrated the flux by our own means. The flux calibration for the GRAVITY and MATISSE data was done with the following steps, for each of the four telescopes: firstly, the data were corrected for the offsets of the absolute wavelength calibration, by checking a sky model for the whole $K$ band using the SkyCalc Model calculator \citep{2012A&A...543A..92N,2013A&A...560A..91J}. We used telluric lines at $\lambda=3.903\,\mu$m for the $L$ and $M$ bands and $\lambda=9.576\,\mu$m for the $N$ band. Table~\ref{tab:offset} shows the offsets of the absolute wavelength calibration for each band and target. 

\begin{table}
\caption{Wavelength offset in $\mu$m based on telluric lines for each target and band. The spectral resolution for the $K$ band is $R\sim4000$, for the $L$ and $M$ bands is $R\sim500$. For the $N$ band it is $R\sim1000$ for AH~Sco and KW~Sgr and $R\sim30$ for the other targets.}
\label{tab:offset}
\small
\centering
\begin{tabular}{c c c c }
    \hline \hline
      Target & $K$ ($\mu$m) & $L/M$ ($\mu$m) & $N$ ($\mu$m) \\
     \hline %\hline
       AH Sco & 0.0003 & 0.0013 & 0.0135 \\
       KW Sgr & - & 0.0002 & 0.0148\\
       V602 Car & 0.0006 & 0.0013 & 0.0403 \\
       CK Car & 0.0006 & 0.0014 & 0.0403 \\
       V460 Car & 0.0006 & 0.0013 & 0.0405 \\

       %VAR Cet & 0.0006 & 0.0002 & 0.0053 \\
 %          \hline
%      ups Cet &  &  $\pm$ & 3800 & 1.15 & -0.05 \\
     \hline   
\end{tabular}

\end{table}

Secondly, the calibrator flux was divided by the MARCS model flux, using the parameters in Table~\ref{tab:calib} (for this step, the MARCS flux was binned to match the resolution of the data) to obtain the spectral transfer function. Finally, the science flux was divided by the spectral transfer function and normalised. 

\medskip
%\subsection{Archival data}
 For KW Sgr, the requested $K$-band VLTI-GRAVITY data could not be obtained, and we used instead published VLTI/AMBER available in \citet{2013A&A...554A..76A}. The data were taken using the AMBER medium-resolution mode R$\sim$1500 in the $K-2.1\,\mu$m and $K-2.3\,\mu$m bands. The reader is referred to \citet{2013A&A...554A..76A} for information about the data acquisition and reduction. 

\subsection{Observational results}\label{sub:ud}
In order to illustrate our data, we computed best-fit uniform disk (UD) diameters to our visibility data as a function of wavelength across the $K$, $L,$ $M,$ and $N$ bands. This fit provides a first estimate of their characteristic sizes in near-continuum, molecular, and dust layers. %The observed flux and visibility data compared to our models of a MARCS model atmosphere with a wind model are discussed in Sect.~\ref{sec:results}.

\begin{figure}
      \centering
      \includegraphics[width=1.\linewidth]{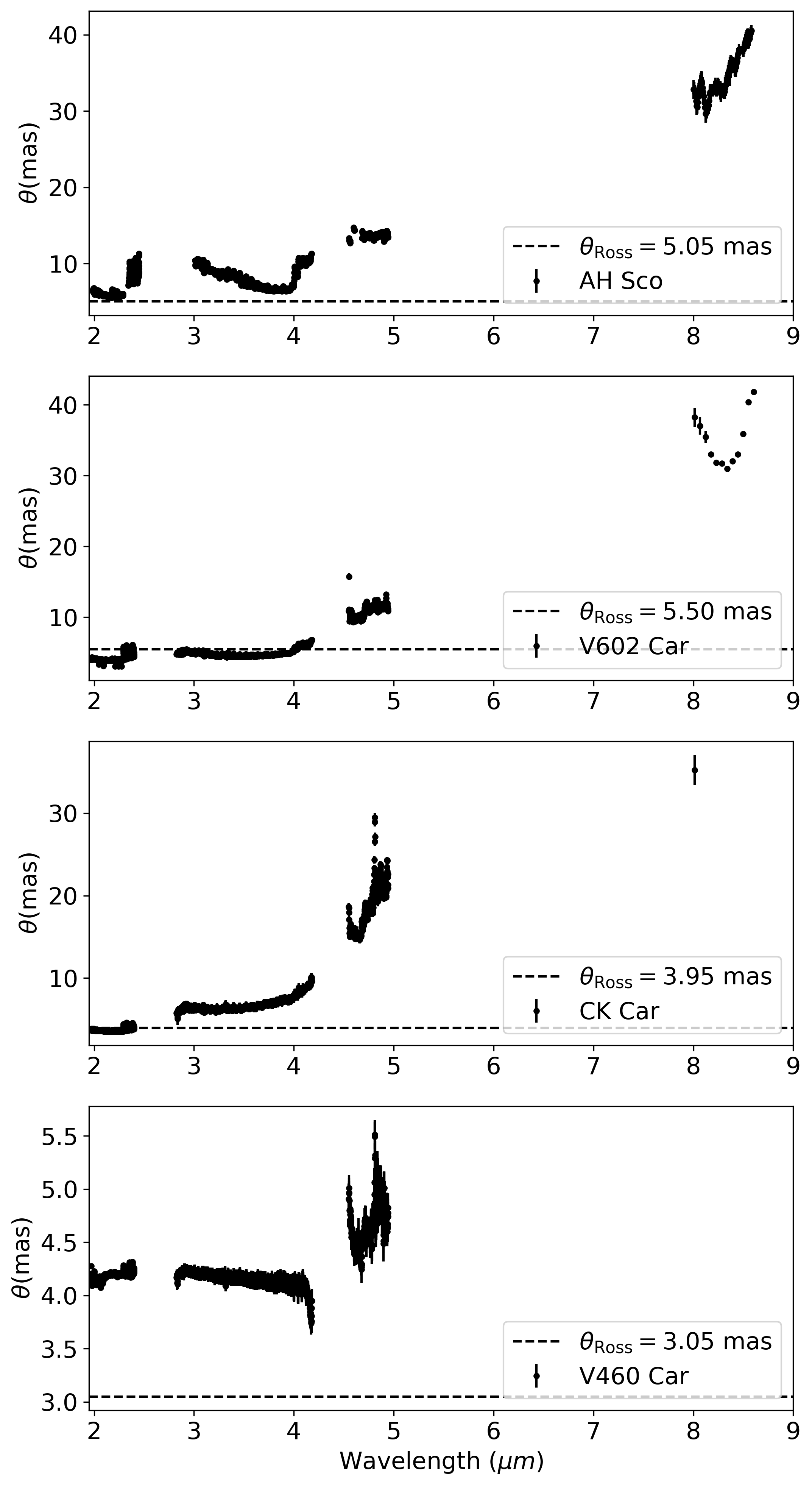}
      \caption{Results for the UD model fit using the AH~Sco, V602~Car, CK~Car, and V460~Car data in the $K$, $L$, $M$, and $N$ bands. Also shown is the Rosseland angular diameter, $\theta_\mathrm{Ross}$, as derived by the model fits described in Sect.~\ref{sub:mod}.
       }
      \label{fig:ud1}
  \end{figure}

Figure~\ref{fig:ud1} shows the resulting UD diameters $\theta_\mathrm{UD}$ for our sources AH Sco, V602 Car, CK Car, and V460Car, respectively, which are the sources for which we have both MATISSE and GRAVITY data. Best-fit Rosseland-mean angular diameters, as obtained from our model fit described in Sect.~\ref{sub:mod} below, are shown as well.
We see that $\theta_\mathrm{UD}$ increases for specific wavelengths where molecules are present. Most notably, 
$\theta_\mathrm{UD}$ increases at the CO band heads at $\lambda>2.3\,\mu$m and at the SiO line at $\lambda=4.0\,\mu$m by about a factor of 2, and in the $M$ band, dominated by CO lines, up to a factor of 3--5. 
In the $N$ band, the targets are too resolved to obtain a reliable UD fit for most data points. This indicates that these molecules are formed in extended atmospheric regions. V460~Car shows clearly less extended layers compared to the other targets,  which could be caused by a lower luminosity and lower mass-loss rate. %\textbf{Note that here we are not assuming any mass loss rate in our model, since we are using a simple UD model fit where the intensity of the UD model, $I_{\mathrm{UD}}$ is described as:
%\begin{equation}\label{eq:ud}
   % I_{\mathrm{UD}} (\rho) =
    %\begin{cases}
    %  4/(\pi\theta^{2}) & \mathrm{if}\;\rho\leq \theta/2\\
    %  0                 & \mathrm{if}\;\rho>\theta/2
   %\end{cases}
%\end{equation}
%where $\rho$ and $\theta$ are the polar coordinates in the object plane \citep{2007NewAR..51..576B}.}
In the region between 8\,$\mu$m and 9\,$\mu$m, $\theta_\mathrm{UD}$ increases by up to a factor of 8--10 relative to the near-continuum bands.

The simple UD fit already illustrates the extended atmosphere of RSGs over a large wavelength range from $1.8<\lambda<12.0\,\mu$m. With our MARCS+wind model discussed in Sect.~\ref{sec:results} we are able to reproduce these extensions for the first time across the $K$, $L$, and $M$ bands. The $N$ band is strongly affected by dust layers that are not included in our model of the extended atmosphere, and as a result, is not discussed further.
 
 \section{Methods}\label{sec:methods}
 
 \subsection{Model}\label{sub:mod}
  The model is based on work by \citet{2021MNRAS.508.5757D} and Paper I. Here we briefly summarise the most important features.
For a {\sc MARCS} model atmosphere, we mimicked the effect of a stellar wind using the mass continuity expression
\begin{equation}\label{eq:masscont}
\dot{M}=4\pi r^{2}\rho(r)v(r)
,\end{equation}
where $\rho$ and $v$ are the density and velocity as a function of the stellar radial coordinate $r$, respectively. The wind density $\rho_{\mathrm{wind}}(r)$ has the following shape, proposed by \citet{2001ApJ...551.1073H}: 
\begin{equation}\label{eq:betalaw}
\rho_{\mathrm{wind}}=\frac{\rho_{\mathrm{phot.}}}{(R_{\mathrm{max}}/R{\star})^{2}} \left( 1-\left( \frac{0.998}{(R_{\mathrm{max}}/R_{\star})} \right) ^{\gamma} \right) ^{\beta}
,\end{equation}
where $R_{\mathrm{max}}$ is the arbitrary outer-most radius of the model, in our case $\sim8.5\,R_{\star}$.  

The $\beta$ and $\gamma$ parameters define the smoothness of the extended wind region. We performed the same analysis as in Paper I: using a grid with $-1.1<\beta<-1.60$ in steps of $\Delta\beta=0.25$ and $0.05<\gamma<0.45$ in steps of $\Delta\gamma=0.2$, our best fit parameters were found to be $\beta=-1.60$ and $\gamma=0.05$ for all sources. These parameters are consistent with the best fit from Paper I.
The velocity profile was found assuming a fiducial wind limit of $v_{\infty}=25\pm5$ km/s, which is the value matched to
 \citet{1998IrAJ...25....7R}, \citet{2005A&A...438..273V}, and \citet{10.1093/mnras/stx3174} and derived from direct measurements and Eq.~\ref{eq:masscont}. 
 
 For the temperature profile, we assumed the same two profiles as in Paper I, one following simple radiative equilibrium (RE) and another assuming a temperature inversion in the chromosphere of the star that peaks at $\sim 1.4\,R_{\star}$ and decreases again, following \citet{2001ApJ...551.1073H}. As in Paper I, the chromospheric temperature profile did not fit well the spectra and $|V|^{2}$ simultaneously, unlike RE. The reason is that this temperature inversion profile shows the CO flux in the region $2.3<\lambda<2.4\,\mu$m as emission lines. Since we do not observe any emission lines in the observations, we  decided to use only the results based on RE. For the example of AH Sco, a fit with both profiles is shown in Appendix~\ref{app:ap1}. A full discussion about the different temperature profiles of the {\sc MARCS}+wind model can be found in Paper I.  
 
 We re-sampled the temperature, density and velocity profiles to a constant logarithmic optical depth sampling $\Delta \log(\tau)$ (see the reasons for this re-sampling in \citet{2021MNRAS.508.5757D} and Paper I). Finally, we defined the outer boundary of the model where the local temperature is $>800$ K. Below this temperature, our code is unable to reliably converge the molecular equilibrium. In addition, some species would be depleted to dust grains. As a clarification, the $\tau_\lambda=1$ surface is always within this radius if $\log \dot{M}/M_{\odot}\mathrm{yr}^{-1}<-4$.

Adding a wind density and temperature profile to the photosphere reproduces simultaneously the observed diameter variations and the spectral line depths on a large wavelength range in the $K$, $L$, and $M$ bands. We cannot exclude the possibility that RSG atmospheres can be extended without assuming a wind (e.g. an extension due to Alfv{\'e}n waves, or convective plumes). However, these limited extensions near the photosphere will not affect the apparent diameter and spectrum to the same extend as a stellar wind. If the mechanism initially lifting material is convection, it may generate asymmetric extension. However, to lift material out of the gravitational well, with an escape velocity of the order of $100\,{\rm km/s}$ at the photosphere,
 requires some additional factor, like radiative pressure on molecular lines \mbox{\citep{2007A&A...469..671J}} and dust. Our assumption of spherical symmetry remains however reasonable as shown by \citet{2002A&A...386.1009P} in the case of Betelgeuse.
 As the scope of this paper is not to discuss the physical processes that extend the atmosphere of RSGs, our model assumes the presence of a spherically symmetric wind without specifying its physical driver.

 \subsection{Analysis}\label{sub:analysis}
 We computed the spectra and intensity profiles using the radiative transfer code \textsc{Turbospectrum v19.1} \citep{2012ascl.soft05004P}, setting a wavelength range from $1.8\,\mathrm{\mu m}$ to $5.0\,\mathrm{\mu m}$ with a step of $0.1 \AA$ in order to resolve the microturbulence. 
 
For the spectral synthesis, we included a list of atomic and molecular data\footnote{The atomic and molecular data are available at \url{https://github.com/bertrandplez/Turbospectrum2019}}. Chemical equilibrium was solved for 92 atoms and their first two ions, including Fe, Ca, Si, and Ti, and about 600 molecular species, including CO, TiO, H$_{2}$O, OH, CN, and SiO.

The stellar parameters assumed in the {\sc MARCS} model for our targets are shown in Table~\ref{tab:marcs}. The parameters were found in \citet{2013A&A...554A..76A,2015A&A...575A..50A} for AH~Sco, KW~Sgr and V602~Car and in \citet{2019MNRAS.490.3158C} for CK Car and V460 Car. 

 \begin{table*}
\caption{ Parameters for the {\sc MARCS} models used for the analysis of each RSG. From left to right: luminosity $\log \mathrm{L}/\mathrm{L}_{\odot}$, effective temperature $T_{\mathrm{eff}}$, surface gravity $\log g$, metallicity [Z], microturbulence $\xi$, and mass as in \citet{2009ApJ...703..420M}, \citet{2013A&A...554A..76A,2015A&A...575A..50A}, and \citet{2019MNRAS.490.3158C}. The last column shows the radius of the star in the photosphere in solar units (defined at $\tau_{\mathrm{Ross}}=2/3$).}
\label{tab:marcs}
\small
\centering
\begin{tabular}{c c c c c c c c}
    \hline \hline
      RSG & $\log \mathrm{L}/\mathrm{L}_{\odot}$ & $T_{\mathrm{eff}}$ (K) & $\log g$ & [Z] & $\xi$ (km/s) & $M/M_{\odot}$ & $R_{\star}/R_{\odot}$ \\
     \hline %\hline
        AH Sco & 5.52 & 3600 & -0.5 & 0 & 5 & 20 & 1411  \\
        KW Sgr & 5.24 & 3700 & 0.0 & 0 & 5 & 20 & 1009  \\
        V602 Car & 5.1 & 3400 & -0.5 & 0 & 5 & 20 & 1015  \\
        CK Car & 4.86 & 3500 & 0.0 & 0 & 5 & 15 & 690  \\
        V460 Car & 4.46 & 3600 & -0.5 & 0 & 5 & 15 & 539  \\ 
        %VAR Cet & 3.95 & 3500 & 0.0 & 0 & 5 & 5 & 256  \\
     \hline
     
\end{tabular}
\end{table*}

 \section{Results}\label{sec:results}
 
We computed the visibilities with the same method as in Paper I. Briefly, we used the Hankel transform defined in \citet{2000MNRAS.318..387D}, estimated the angular diameter of the outermost layer of the model ($\theta_{\mathrm{Model}}$) using the relation with the Rosseland angular diameter $\theta_{\mathrm{Ross}}$ found in \citet{2000MNRAS.318..387D} and \citet{2004A&A...413..711W}, and scaled the final model with an $A$ factor that allows for the attribution of a fraction of the flux to an over-resolved circumstellar component \citep{2013A&A...554A..76A}. This $A$ component depends on the wavelength range of the observation.

To estimate both $\theta_{\mathrm{Ross}}$ and $A$, for each studied model, we computed the $|V|^{2}$ as a function of their spatial frequency $B/\lambda$ in the whole wavelength range of the $K$, $L,$ and $M$ bands. We compared the model and data $|V|^{2}$ of both instruments, and found the best-fit $\theta_{\mathrm{Ross}}$ and $A$ by means of a $\chi^{2}$ minimisation. The $\chi^{2}$ is computed using the mean $|V|^2$ of the following near-continuum regions and fitting to our best model: $2.23-2.27\,\mu$m, $3.25-3.45\,\mu$m, $4.60-4.75\,\mu$m for the $K$, $L,$ and $M$ bands, respectively. For the case of KW~Sgr, we used the AMBER data instead. We found the same best-fit $A_{L}$ and $A_{M}$ values for the $L$ and $M$ bands, while the best-fit $A_{K}$ in the $K$ band was different. 

Table~\ref{tab:theta-a} shows the best fit for $\theta_{\mathrm{Ross}}$, $A_{\mathrm{L/M}}$ and $A_{\mathrm{K}}$ for each of our RSGs. The errors in $\theta_{\mathrm{Ross}}$ and the $A$ values were derived by the minimum values in the 68\% dispersion contours of the $\chi^{2}$ fit, which for 2 degrees of freedom corresponds to $\chi^{2}<\chi_{\mathrm{min}}^{2}+2.3$ \citep{1976ApJ...210..642A}. Our results are in disagreement with \citet{2013A&A...554A..76A,2015A&A...575A..50A} within the error limits, but agree within $3\sigma$. This disagreement could be because we are fitting $\theta_{\mathrm{Ross}}$ for all the bands simultaneously. Previous studies such as \citet{2013A&A...554A..76A,2015A&A...575A..50A} use only the $2.25\,\mu$m near-continuum spectral window.

We used a range of mass-loss rates of $-7<\log \dot{M}/M_{\odot}\mathrm{yr}^{-1}<-4$ with a grid spacing of $\Delta\log \dot{M}/M_{\odot}\mathrm{yr}^{-1}=0.5$ to determine the best-fit $\dot{M}$. In the last column of Table~\ref{tab:theta-a} we also show the best-fit $\log \dot{M}/M_{\odot}\mathrm{yr}^{-1}$ for our static wind model and the errors corresponding to the grid spacing.

 \begin{table*}
\caption{Best-fit interferometric parameters and mass-loss rate for each RSG. From left to right: the best-fit $\theta_{\mathrm{Ross}}$, the $A$ value for the $K$ and $L/M$ bands, respectively, and the $\log \dot{M}/M_{\odot}\mathrm{yr}^{-1}$ fit from the model. The fixed parameters of our model for all targets are the density profile parameters $\beta=-1.60$ and $\gamma=0.05$, and simple RE for the temperature profile.}
\label{tab:theta-a}
\small
\centering
\begin{tabular}{c c c c c}
    \hline \hline
      RSG & $\theta_{\mathrm{Ross}}$ (mas) & $A_{\mathrm{L/M}}$ & $A_{\mathrm{K}}$ &  $\log \dot{M}/M_{\odot}\mathrm{yr}^{-1}$ \\
     \hline %\hline
        AH Sco & 5.05 $\pm$ 0.5 & 0.75 $\pm$ 0.05 & 0.85 $\pm$ 0.05 & -4.0 $\pm$ 0.50\\
        KW Sgr & 3.00 $\pm$ 0.4 & 0.85 $\pm$ 0.05 & 0.9 $\pm$ 0.05 & -4.5 $\pm$ 0.50\\
        V602 Car & 5.5 $\pm$ 0.5 & 0.7 $\pm$ 0.05 & 0.85 $\pm$ 0.05 & -5.0 $\pm$ 0.50\\
        CK Car & 3.95 $\pm$ 0.5 & 1 $\pm$ 0.05 & 1 $\pm$ 0.05 & -5.0 $\pm$ 0.50\\
        V460 Car & 3.05 $\pm$ 0.6 & 0.95 $\pm$ 0.05 & 0.95 $\pm$ 0.05 &  -6.5 $\pm$ 0.50\\
        %VAR Cet & 4.65 $\pm$ 0.6 & 0.65 $\pm$ 0.05 & 0.90 $\pm$ 0.05 & -6.5 $\pm$ 0.50\\
     \hline
     
\end{tabular}
\end{table*}

Checking the $\log \dot{M}/M_{\odot}\mathrm{yr}^{-1}$ column in Table~\ref{tab:theta-a}, our best-fit mass-loss rate is in accordance with typical mass-loss prescriptions \citep[e.g.][]{1988A&AS...72..259D,2005ApJ...630L..73S}. As the stars become more luminous, the $\dot{M}$ becomes very high compared to recent prescriptions \citep[e.g.][]{2020MNRAS.492.5994B}, which is the case of AH~Sco and KW~Sgr. This could mean that our model has still some limitations. Even though we checked that we are not hitting the edge of our model grids, it could be that the temperature or density profiles are still not optimum. Another reason for this high $\dot{M}$ would be that since the pure MARCS model is very compact, we could need an unrealistically higher $\dot{M}$ to reproduce the observations.

\subsection{GRAVITY}

\begin{figure}
      \centering
      \includegraphics[width=.9\linewidth]{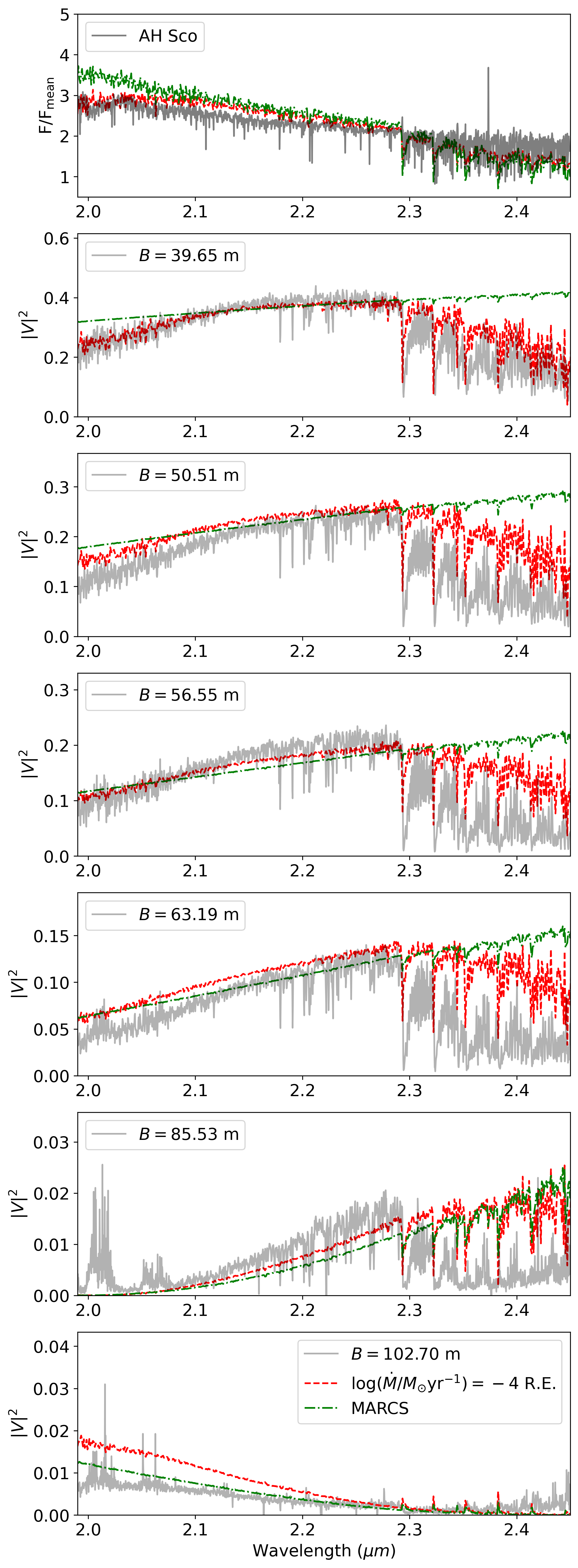}
      \caption{GRAVITY data, best-fit model, and pure {\sc MARCS} model fit for AH~Sco. \textit{Upper panel: } Normalised flux for the RSG AH~Sco (grey), as observed with VLTI/GRAVITY for the $K$ band. Our best-fit model is shown in red for RE, while the pure {\sc MARCS} model fit is shown in green. \textit{Lower panels: } Same as the upper panel but for the $|V|^{2}$ with different baselines. Both flux and $|V|^{2}$ are better represented by our fit. }
      \label{fig:ahscog}
  \end{figure}

Figures~\ref{fig:ahscog}, \ref{fig:v602carg}, \ref{fig:CKcarg}, and \ref{fig:v460carg} show the reduced data for the $K$ band, the best-fit {\sc MARCS}+wind model for the temperature profile of RE and the initial best pure {\sc MARCS} model fit for AH~Sco, V602~Car, CK~Car, and V460~Car, respectively. We include every baseline of the observations. Our RE model reproduces better both the fluxes and $|V|^{2}$ than the pure {\sc MARCS} model. 

The most important features that our model can reproduce in this wavelength range are the CO absorption lines at $\lambda=2.29-2.5\,\mathrm{\mu m}$ as well as the water signatures at $\lambda=1.8-2.2\,\mathrm{\mu m}$ and towards the upper edge of the $K$ band at $\lambda>2.35\,\mathrm{\mu m}$. These features are present as a broader shape in the absorption lines in the flux as well as by lower $|V|^{2}$ values. We see that for all the cases, while the pure MARCS model matches our model and the data in the spectra, the $|V|^{2}$ extensions for the different baselines are substantially better reproduced by our model as compared to pure MARCS, specifically for CO. This means that the layers where the CO lines are formed are more extended than in the MARCS models.

    \subsection{MATISSE}
\begin{figure}
      \centering
      \includegraphics[width=1.\linewidth]{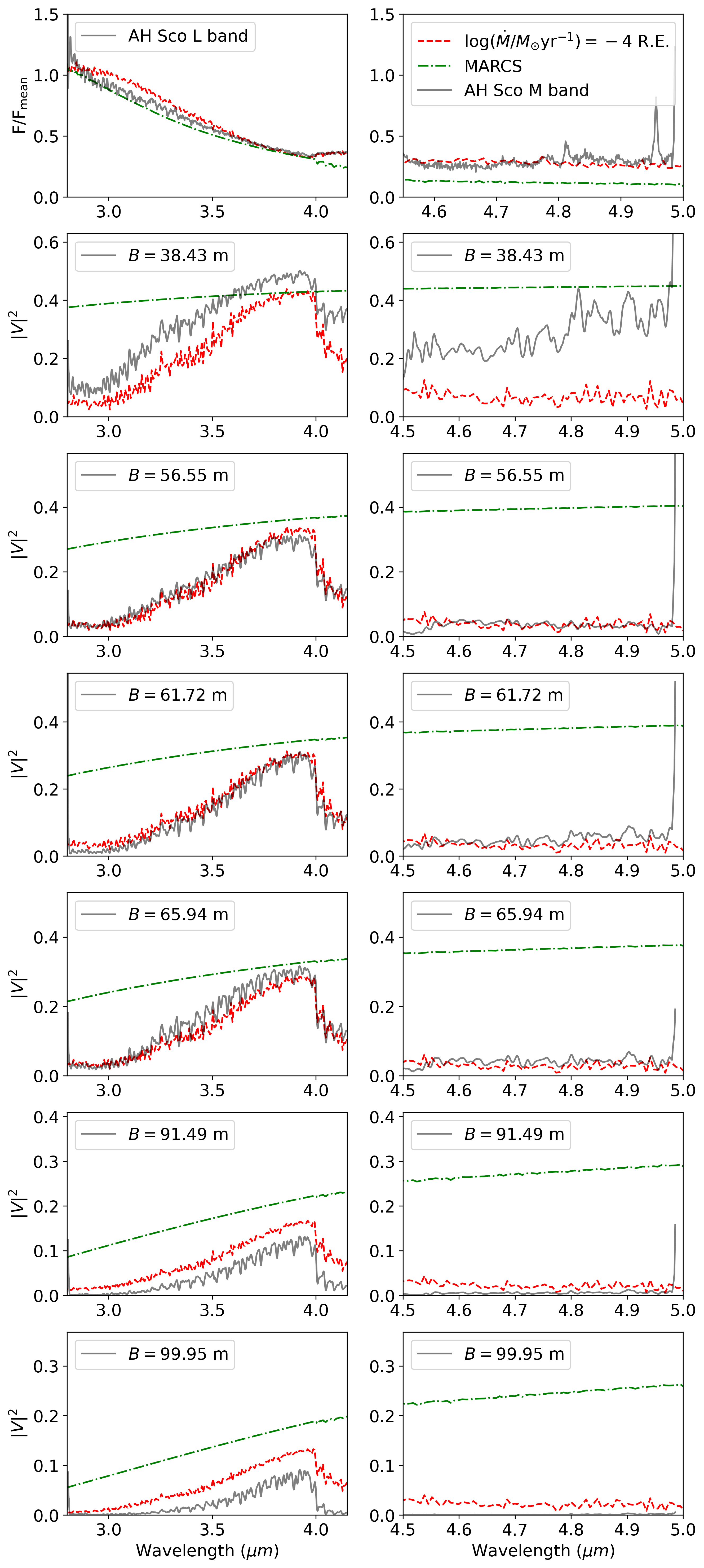}
      \caption{MATISSE data, best-fit model, and pure {\sc MARCS} model fit for AH~Sco. \textit{Upper panel: } Normalised flux for the RSG AH~Sco (grey), as observed with VLTI/MATISSE for the $L$ (\textit{left panels}) and $M$ bands (\textit{right panels}). Our best-fit model is shown in red, while the pure {\sc MARCS} model fit is shown in green. \textit{Lower panels: } Same as the upper panel but for the $|V|^{2}$ with different baselines. Both flux and $|V|^{2}$ are better represented by our fit. }
      \label{fig:ahscom}
  \end{figure}

  Figures~\ref{fig:ahscom}, \ref{fig:kwsgrm}, \ref{fig:v602carm}, \ref{fig:CKcarm}, and \ref{fig:v460carm} show the MATISSE reduced data for the $L$ and $M$ bands, the best-fit {\sc MARCS}+wind model for the temperature profiles with RE and the initial best pure {\sc MARCS} model fit for AH~Sco, KW~Sgr, V602~Car, CK~Car, and V460~Car, respectively. Our model reproduces better both the fluxes and $|V|^{2}$ than the pure {\sc MARCS} model. 

  The most important feature that our model can reproduce in the $L$ band is the SiO absorption band at $\lambda=4.0\,\mathrm{\mu m}$. Our model matches better the spectra than MARCS at $\lambda>4.0\,\mu$m, specifically for AH~Sco (Fig.~\ref{fig:ahscom}), where the SiO features appear in emission and MARCS does not reproduce. Regarding the $|V|^{2}$ extensions for the different baselines, pure MARCS does not reproduce at all the extension in SiO, while our model can fit the data and reproduce the SiO feature.

In the $M$ band, our model is consistent with observations, lower than the pure MARCS prediction.
This is due to the presence of CO in this band. In this band, RE and the chromospheric temperature profile have strong differences (see Paper I for an extended discussion), where the chromospheric temperature profile shows CO in emission in this wavelength region. Four our targets, RE still reproduces better this wavelength region. 

  The reduced MATISSE $N$-band data are shown in Appendix~\ref{app:ap2}. It has not been analysed since that wavelength region is highly affected by dust, and therefore cannot be predicted by our model. Checking the shape of the flux and $|V|^{2}$ in Appendix~\ref{app:ap2}, the most prominent molecular feature is SiO around $8.2\,\mu$m, followed by silicates and Al$_{2}$O$_{3}$ dust \citep{2017A&A...600A.136P,2022A&A...658A.185C}.

 \subsection{AMBER}
   Figure~\ref{fig:kwsgra} shows the published AMBER data in the $K$ band by \citet{2013A&A...554A..76A}, the best-fit {\sc MARCS}+wind model and the initial best pure {\sc MARCS}, since we did not have GRAVITY data for KW~Sgr. 

 In Fig.~\ref{fig:kwsgra}, the depth of the CO bands ($\lambda>2.3\,\mu$m) it is not matched by our model, while the same model fits better the MATISSE data (Fig.~\ref{fig:kwsgrm}). As mentioned in Paper I, a possible explanation for this mismatch could be that for increasing $\dot{M}$, the models still fail to reproduce the extension of the water or CO layers. In addition, since our model neglects velocity gradients, it underestimates the equivalent widths of lines, which increases the apparent stellar extension at those wavelengths.%This is also the case for AH~Sco and V602~Car MATISSE data to a lower extent.

%-----------------------------------------------------------------
\section{Discussion: Formation of SiO}\label{sec:diss}

After finding the optimised model for each of the targets, the profiles are consistent with the characteristics found in Paper I: simple RE for the temperature and a density profile with $\beta=-1.60$ and $\gamma=0.05$. Moreover, we see that both a single density and temperature profile fits well both GRAVITY and MATISSE observations simultaneously, as well as the AMBER observation to a certain extent.

%\subsection{Formation of SiO}
    
SiO is one of the most important molecules in the atmosphere and circumstellar envelope of evolved cool massive stars, as it is a precursor of silicate dust formation. MARCS models alone cannot explain the presence of SiO lines in cooler red giants and RSGs even when taking into account the dust emission. This means that the SiO is formed in the extended molecular atmosphere \citep{2014A&A...561A..47O}. 

Mapping the extended molecular atmosphere where the SiO is present is important to understand the mass-loss mechanisms in cool evolved stars. Indeed, the wind acceleration takes place in the region between the photosphere and the inner region of the circumstellar envelope \citep[i.e. $1\lesssim R \lesssim 6\,R_{\star}$, ][]{2014A&A...561A..47O}. Some attempts to understand the dust formation processes and to map the circumstellar environment in cool evolved stars include the \textsc{ATOMIUM} project using the Atacama Large Millimeter/submillimeter Array \citep[ALMA,][]{2022A&A...660A..94G} or more recently, the Very Large Telescope (VLT) Spectro-Polarimetric High-contrast Exoplanet REsearch (SPHERE) observations \citep{2023A&A...671A..96M}.

\begin{figure}
      \centering
      \includegraphics[width=1.\linewidth]{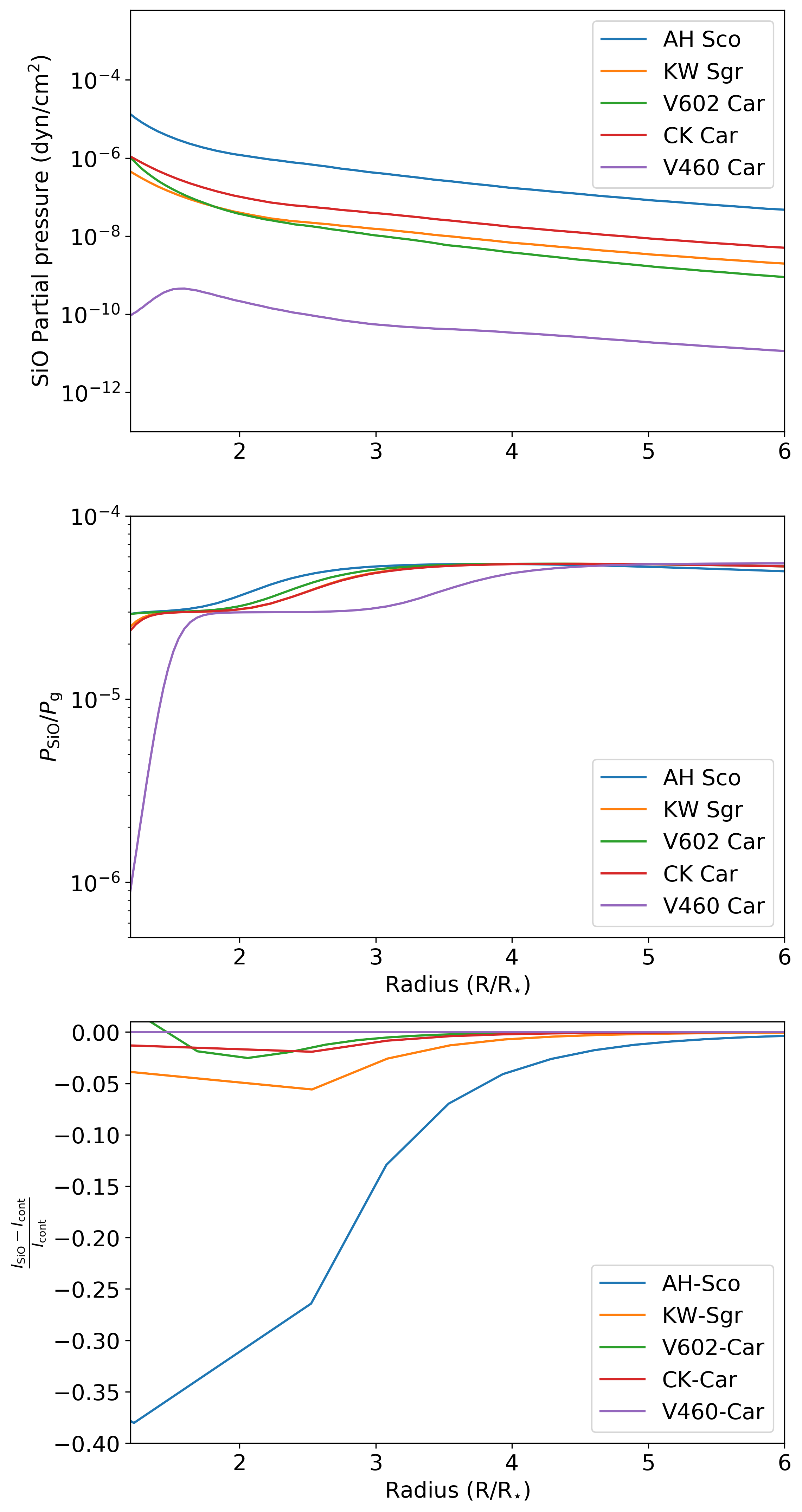}
      \caption{Partial pressure of the SiO, the ratio of the SiO partial pressure, and the gas pressure and intensity at the SiO line minus the continuum  with respect to the stellar radius for each different best-fit model corresponding to the RSG studied. }
      \label{fig:sioh1Dre}
  \end{figure}

Our work focuses on the region where the wind is accelerated. Using our $L$-band MATISSE data, we can study the presence and formation of SiO in this extended atmospheric region. Figure~\ref{fig:sioh1Dre} shows the partial pressure of SiO, this SiO partial pressure divided by the gas pressure ($P_{\mathrm{SiO}}/P_{\mathrm{g}}$) and the relative intensity at the SiO band, defined as the intensity at $\lambda=4.0\,\mu$m minus  the mean intensity at the near continuum $\lambda=3.8\pm0.01\,\mu$m, divided by the mean intensity at the near continuum $\lambda=3.8\pm0.01\,\mu$m ($(I_{\mathrm{SiO}}-I_{\mathrm{cont.}})/I_{\mathrm{cont.}}$), all with respect to the radius of the star for the  best-fit models of our targets. The SiO-band intensity increases at $R\sim2\,R_{\star}$, while the SiO partial pressure decreases and the $P_{\mathrm{SiO}}/P_{\mathrm{g}}$ increases at $R\sim2\,R_{\star}$ until it reaches a constant value at $R\sim3\,R_{\star}$ for all cases except V460 Car. At this radius and beyond, virtually all Si atoms are associated in SiO molecules. The SiO-band intensity decreases, however, as the gas pressure and temperature drop and the optical depth along the line of sight diminishes. We also see in Fig.~\ref{fig:sioh1Dre} that the relative intensity depends on the luminosity used for our models, since the more luminous models have more mass-loss. The gas and SiO pressures are therefore higher, leading to stronger SiO lines. Previous work by \citet{2020A&A...644A.139G} shows that SiO will condense into dust only at larger radii when the temperature has dropped to around $1000\,{\rm K}$.

\section{Summary and conclusion}\label{sec:conc}
We used the 1D modelling approach to add a wind to a pure MARCS model, as in \citet{2021MNRAS.508.5757D} and Paper I, and compared our results with new MATISSE and GRAVITY data of five RSGs. Aside from obtaining the CO and water lines with the GRAVITY data, the MATISSE data add the SiO lines at $\lambda=4\,\mu$m and additional CO transitions in the $M$ band. By comparing the synthetic interferometric visibility and synthetic flux spectra to the data, we see that our models can reproduce the $|V|^{2}$ for the observations in all baselines better than the pure MARCS model. We simultaneously fitted the MATISSE and GRAVITY wavelength coverage. Therefore, we were able, for the first time, to reproduce extensions in the $L$ and $M$ bands, as well as reproduce the extensions for the $K$ band, as in Paper I. 

Our best-fit mass-loss rate increases as the star becomes more luminous. The density and temperature stratification profiles of our model are in accordance with the profiles found in Paper I: simple RE for the temperature profile and a steeper density profile with smoothness parameters of $\beta=-1.60$ and $\gamma=0.05$ (as defined in Eq.~\ref{eq:betalaw}).

Both the partial pressure of SiO relative to the gas pressure, $P_\mathrm{SiO}/P_\mathrm{g}$, and the SiO 4.0\,$\mu$m line intensity increase between 2 and 3 stellar radii. The relative intensity in the SiO line depends on the luminosity used for our models, since the more luminous models have a higher mass-loss rate.

This is one of the first studies of RSGs to use the MATISSE instrument after \citet{2022A&A...658A.185C}. Moreover, \citet{2022A&A...658A.185C} used low spectral resolution, while we obtained our data with medium spectral resolution. Complementing our data with GRAVITY observations in the $K$ band, we have obtained some of the most complete spectro-interferometric datasets to date of RSGs, covering the wavelength range $1.8<\lambda<13\,\mu$m. 

To conclude, this paper shows the potential of the simple 1D MARCS+wind model to fit several spectral bands, providing accurate extensions in the $|V|^{2}$ for the first time. This model approach could be adapted to other lower- and intermediate-mass stars, such as Miras. In the future, we intend to compare this result with recent 3D extended models \citep[e.g.][]{2023A&A...669A..49A}.

\begin{acknowledgements}
We would like to thank the anonymous referee for their useful comments which helped to improve paper. 
Based on observations collected at the European Southern Observatory under ESO programmes IDs 109.231U.002, 110.23P1.001, 0101.D-0616(B) and 110.23.P1.002. GGT has been supported by an ESO studentship and a scholarship from the Liverpool John Moores University (LJMU). GGT is currently supported by the German Deutsche Forschungsgemeinschaft (DFG) under Project-ID 445674056 (Emmy Noether Research Group SA4064/1-1, PI Sander) and Project-ID 496854903 (SA4064/2-1, PI Sander).
\end{acknowledgements}

   \bibliographystyle{aa} % style aa.bst
   \bibliography{sample} % your references Yourfile.bib

\begin{appendix}
%\appendix
\section{GRAVITY, MATISSE, and AMBER observations and best-fit models}\label{app:ap0}
  \begin{figure}
      \centering
      \includegraphics[width=.93\linewidth]{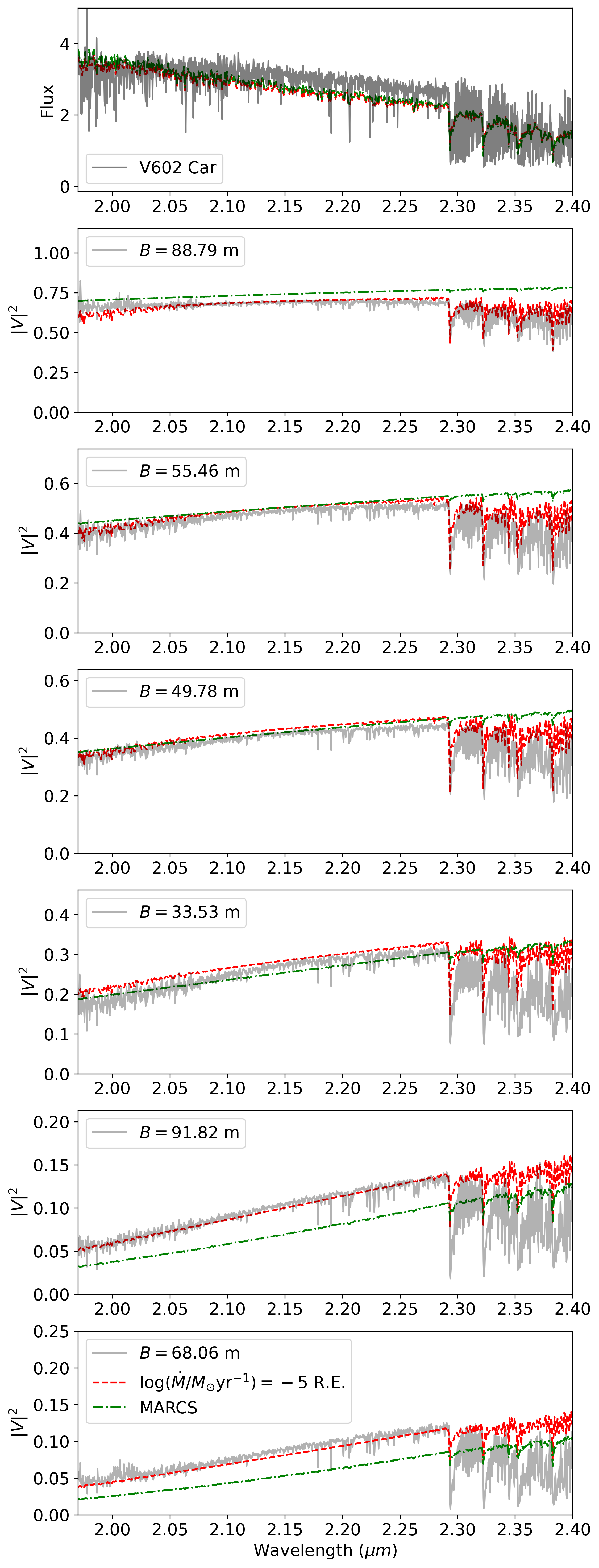}
       \caption{Same as Fig.~\ref{fig:ahscog} but for V602~Car. }
      \label{fig:v602carg}
  \end{figure}

      \begin{figure}
      \centering
      \includegraphics[width=.93\linewidth]{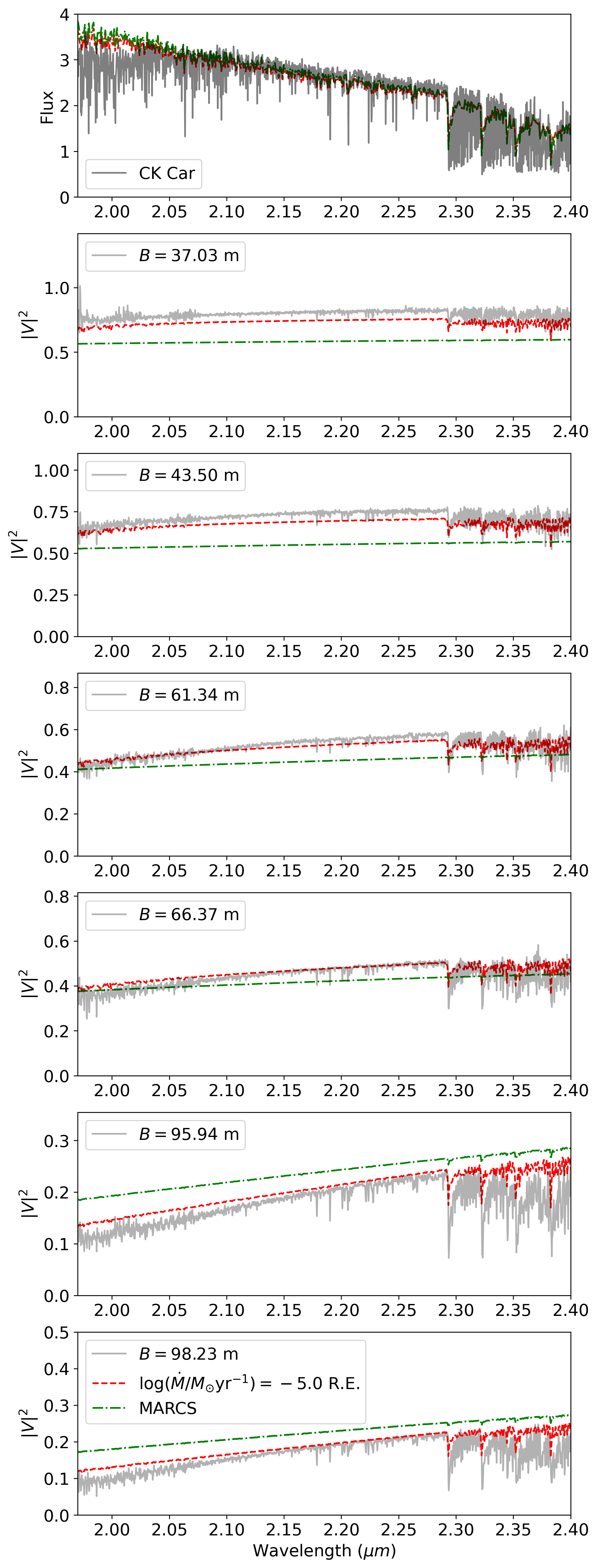}
      \caption{Same as Fig.~\ref{fig:ahscog} but for CK~Car.}
      \label{fig:CKcarg}
  \end{figure}

        \begin{figure}
      \centering
      \includegraphics[width=.93\linewidth]{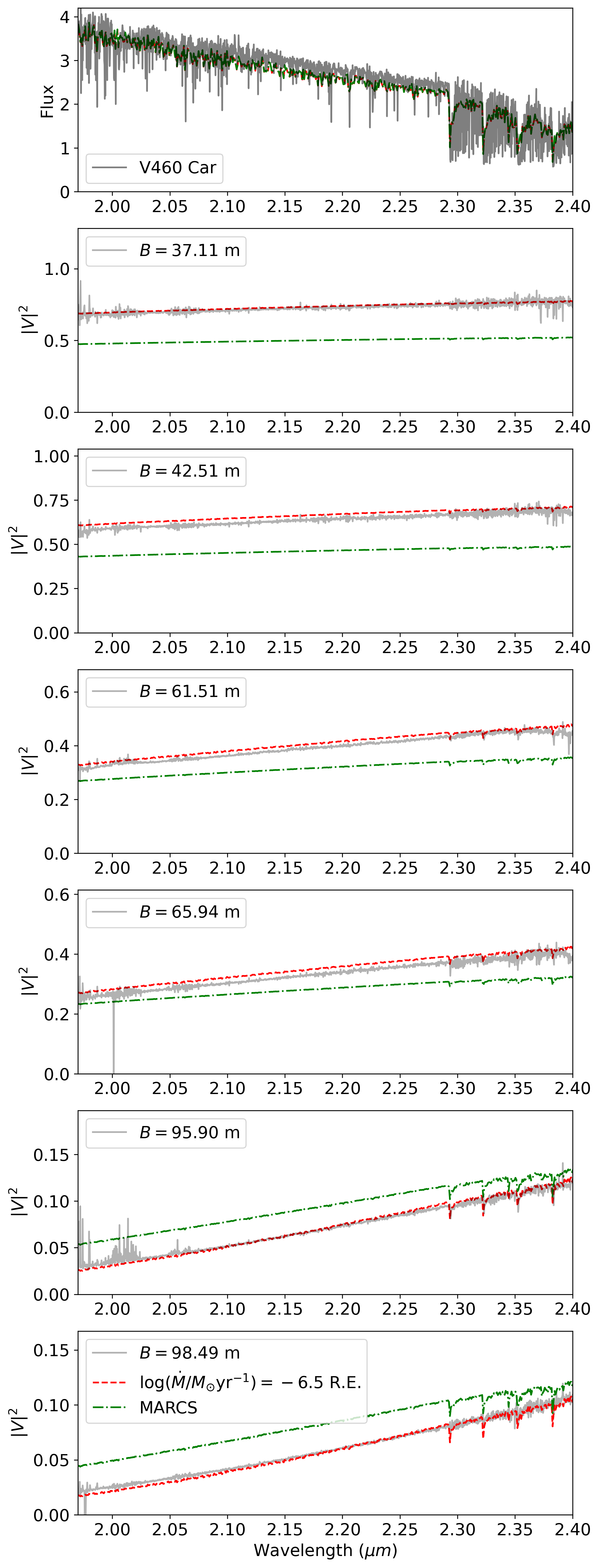}
      \caption{Same as Fig.~\ref{fig:ahscog} but for V460~Car.}
      \label{fig:v460carg}
  \end{figure}
  
    \begin{figure}
      \centering
      \includegraphics[width=1.\linewidth]{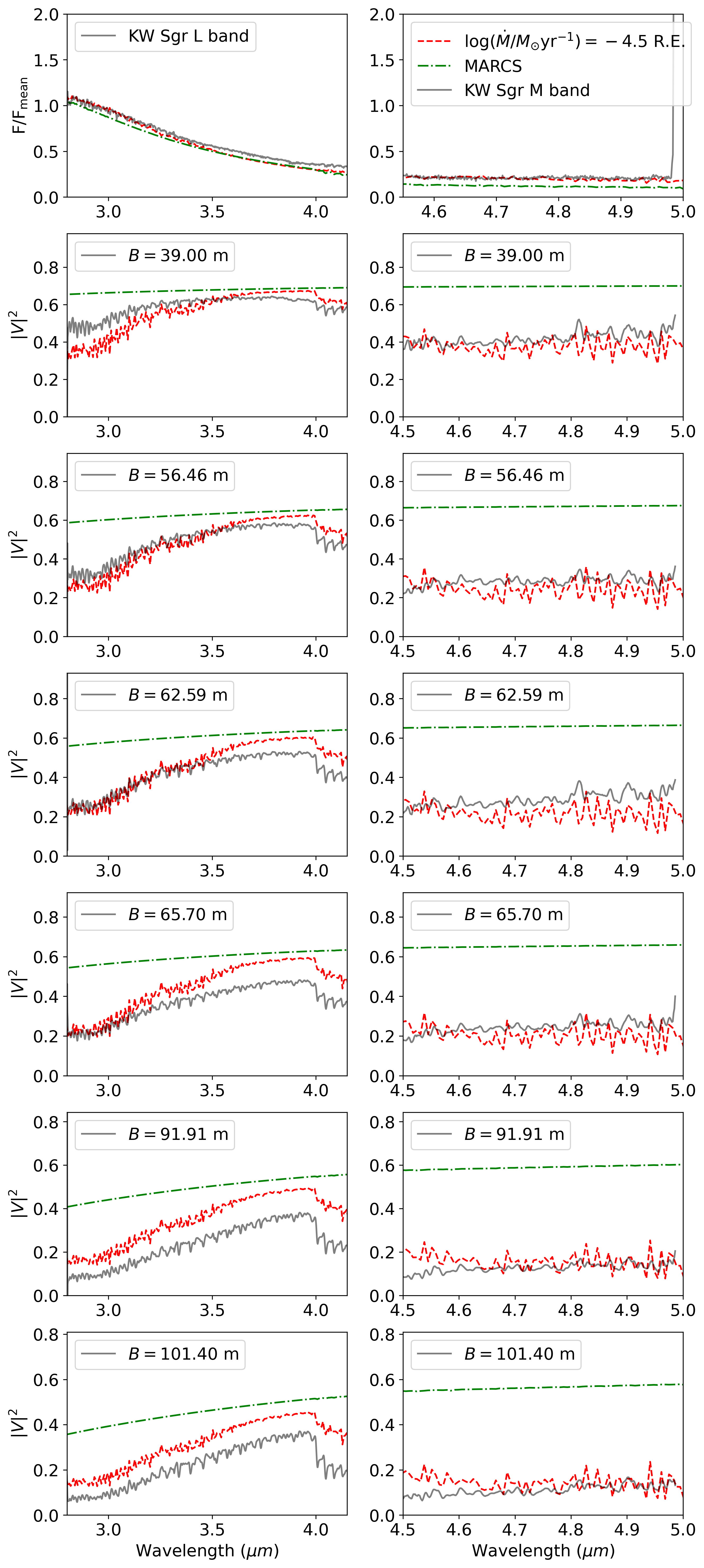}
      \caption{Same as Fig.~\ref{fig:ahscom} but for KW~Sgr.}
      \label{fig:kwsgrm}
  \end{figure}

             \begin{figure}
      \centering
      \includegraphics[width=1.\linewidth]{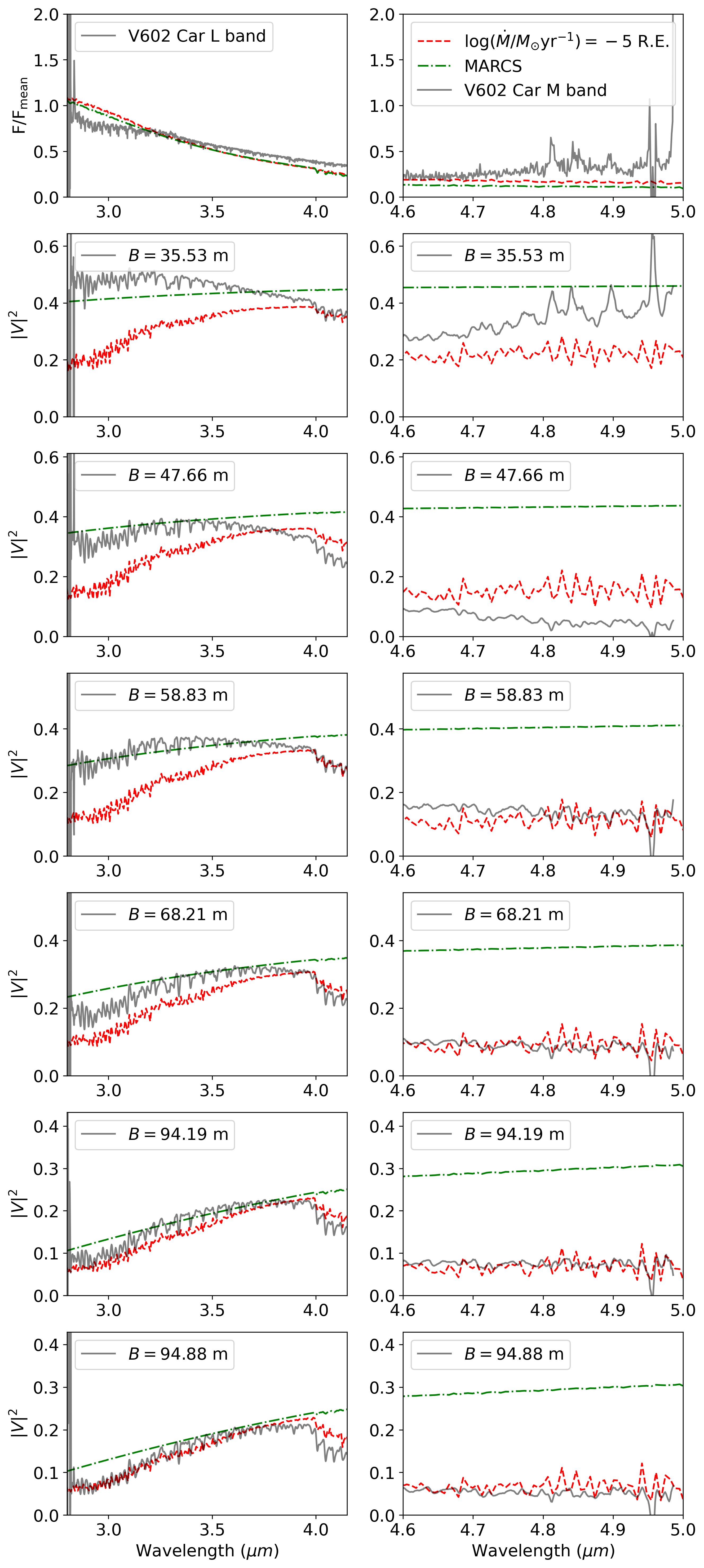}
      \caption{Same as Fig.~\ref{fig:ahscom} but for V602~Car.}
      \label{fig:v602carm}
  \end{figure}

             \begin{figure}
      \centering
      \includegraphics[width=1.\linewidth]{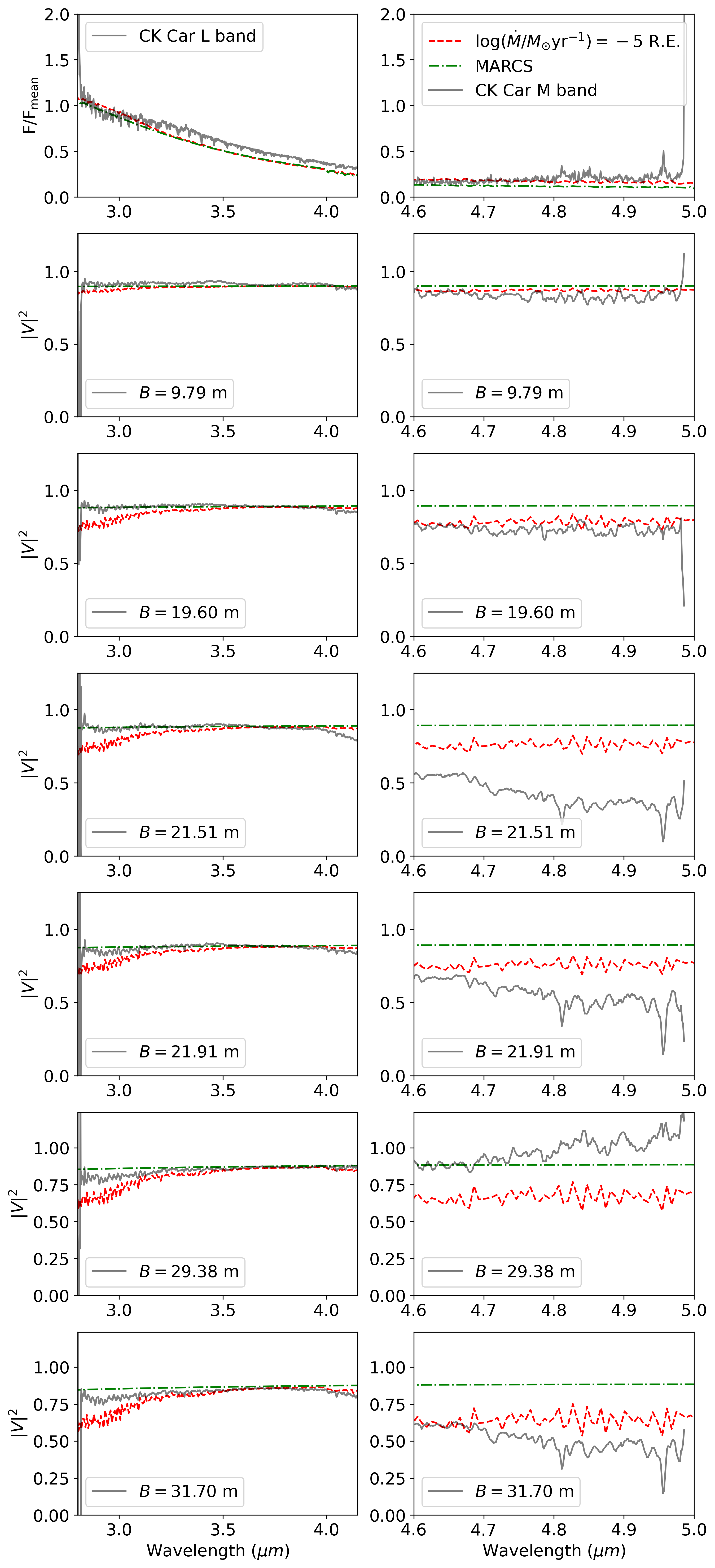}
      \caption{Same as Fig.~\ref{fig:ahscom} but for CK~Car.}
      \label{fig:CKcarm}
  \end{figure}

              \begin{figure}
      \centering
      \includegraphics[width=1.\linewidth]{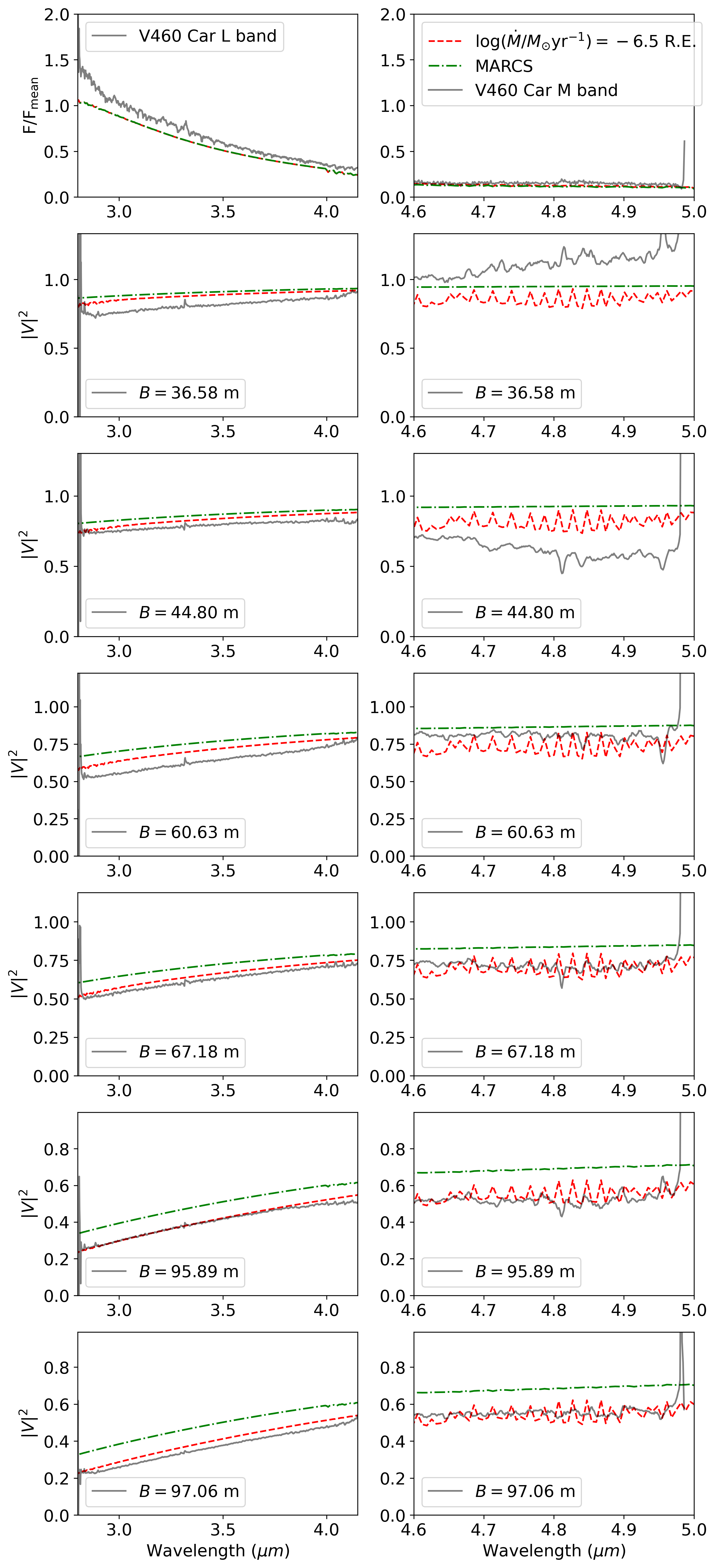}
      \caption{Same as Fig.~\ref{fig:ahscom} but for V460~Car.}
      \label{fig:v460carm}
  \end{figure}
  
        \begin{figure*}
      \centering
      \includegraphics[width=.9\linewidth]{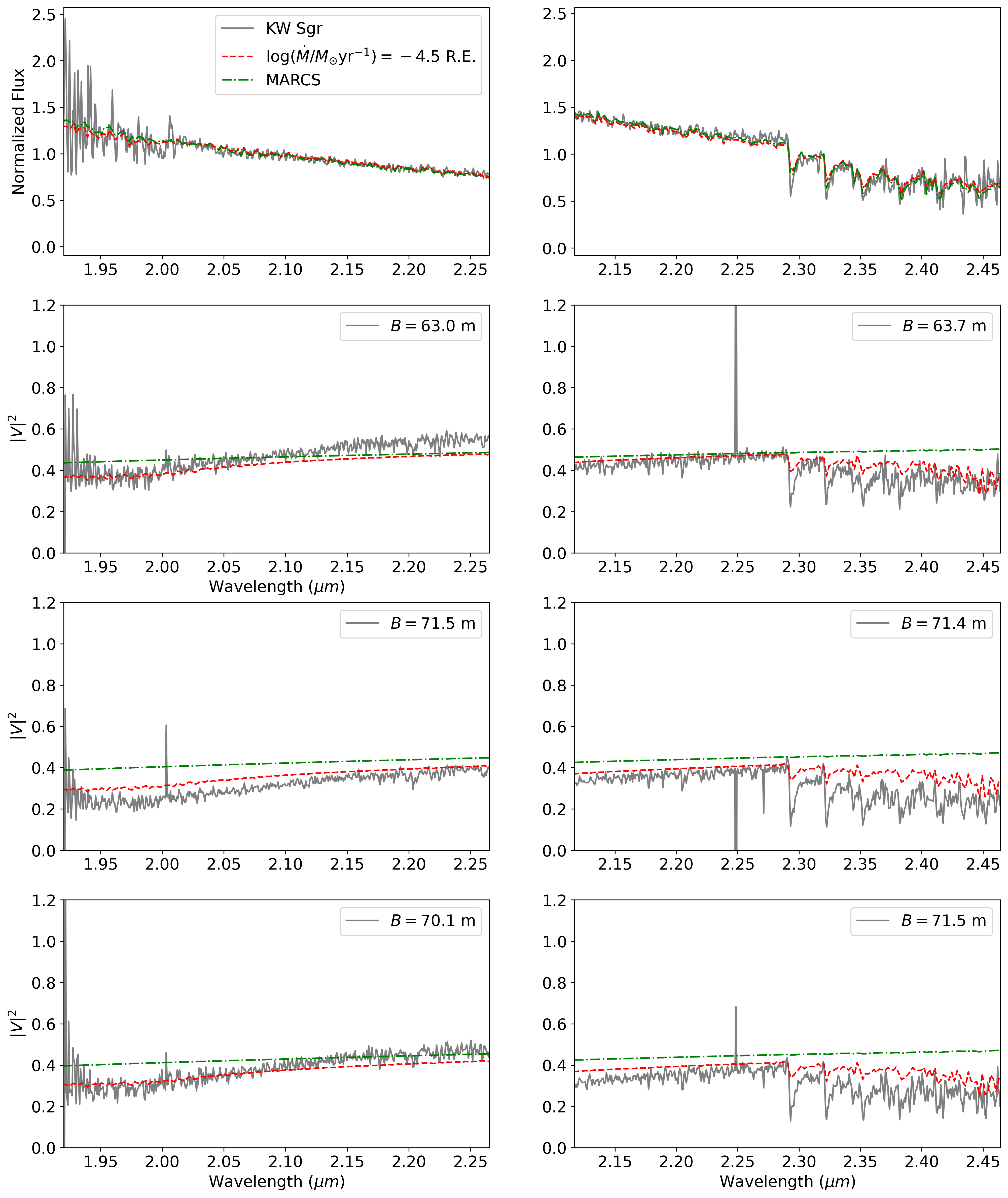}
      \caption{AMBER data, best-fit model, and pure {\sc MARCS} model fit for KW~Sgr. \textit{Upper left: } Normalised flux for the RSG KW~Sgr (grey), as observed with VLT/AMBER for the $K-2.1\,\mathrm{\mu m}$ bands. Our best-fit model is shown in red, and the pure {\sc MARCS} model fit is shown in green. As expected, the fluxes are well represented by both our fit and {\sc MARCS}. \textit{Upper right: } Same as the upper-left panel, but for the $K-2.3\,\mathrm{\mu m}$ band. \textit{Lower left: } Same as the upper-left panel, but for the $|V|^{2}$ with the different baselines. \textit{ Lower right:} Same as the lower-right panel, but for the $K-2.3\,\mathrm{\mu m}$ band and different baselines. Our model can represent the data better than a pure {\sc MARCS} model.}
      \label{fig:kwsgra}
  \end{figure*}
  
%\section{SiO profiles}\label{app:ap00}
\section{Chromospheric temperature profile fit}\label{app:ap1}
      \begin{figure}
      \centering
      \includegraphics[width=.9\linewidth]{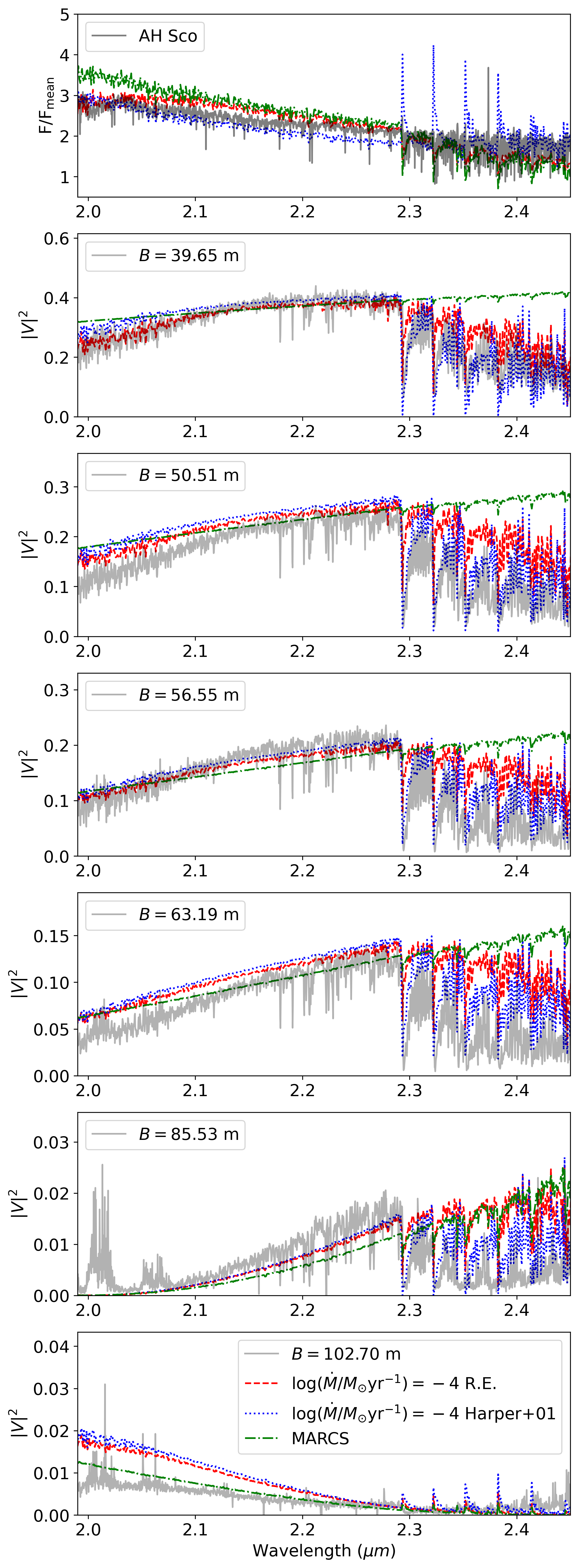}
      \caption{GRAVITY data, best-fit model with both temperature profiles as defined in Paper I, and pure {\sc MARCS} for AH~Sco. \textit{Top panel:} Normalised flux for the RSG AH~Sco (grey), as observed with VLTI/GRAVITY in the $K$ band. Our best-fit model is shown in red for RE and in blue for the chromospheric temperature model, while the pure {\sc MARCS} model fit is shown in green. \textit{Remaining panels:} Same as the upper panel but for the $|V|^{2}$ with different baselines (B). We see that the CO lines are in emission for the chromospheric temperature profile flux, which is not observed in the data. }
      \label{fig:ahscogh}
  \end{figure}

              \begin{figure}
      \centering
      \includegraphics[width=1.\linewidth]{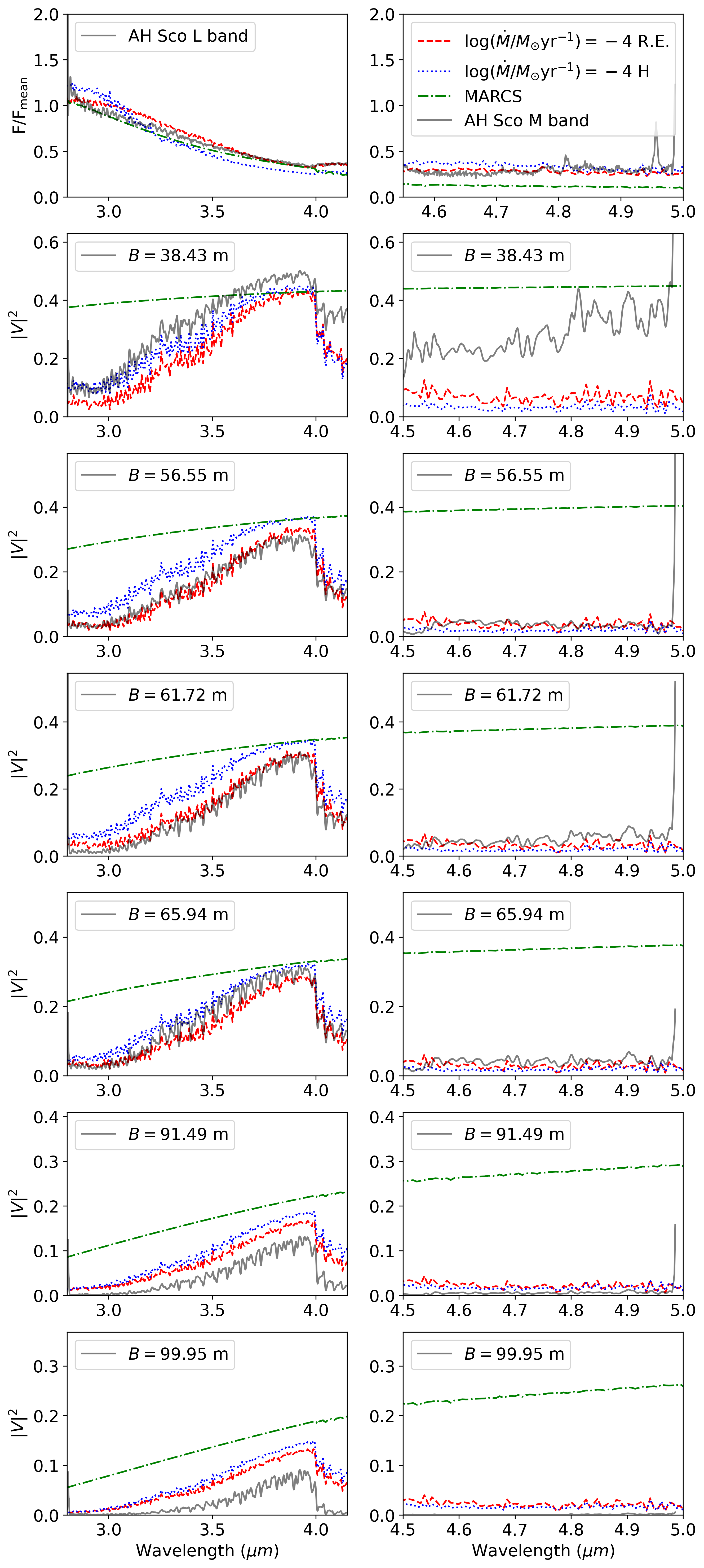}
      \caption{MATISSE data, best-fit model with both temperature profiles as defined in Paper I, and pure {\sc MARCS} for AH~Sco. \textit{Top panels:} Normalised flux for the RSG AH~Sco (grey), as observed with VLTI/MATISSE in the $L$ (\textit{right panels}) and $M$ bands (\textit{left panels}). Our best-fit model is shown in red and the chromospheric temperature profile in blue, while the pure {\sc MARCS} model fit is shown in green. \textit{Lower panels:} Same as the upper panel but for the $|V|^{2}$ with different baselines. Again, both flux and $|V|^{2}$ are better represented by our RE fit. }
      \label{fig:ahscomh}
  \end{figure}

\section{VLTI/MATISSE N-band data}\label{app:ap2}
\begin{figure}
      \centering
      \includegraphics[width=1.\linewidth]{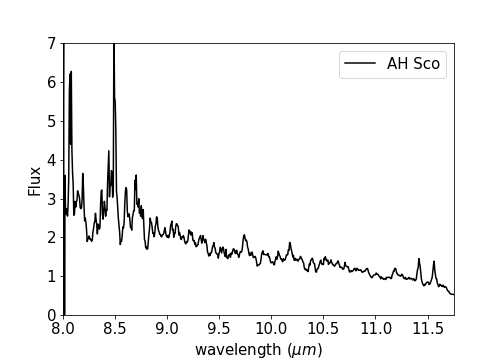}
      \caption{Reduced $N$-band normalised flux for AH~Sco. We excluded the region $\lambda>12\,\mu$m because it was polluted by very high noise. }
      \label{fig:ahfn}
  \end{figure}
  \begin{figure}
      \centering
      \includegraphics[width=1.\linewidth]{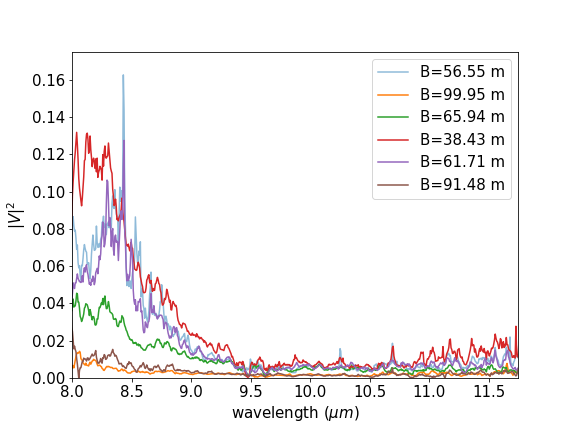}
      \caption{Reduced $N$-band $|V|^{2}$ for AH~Sco, for each of the four beams. We excluded the region $\lambda>12\,\mu$m because it was polluted by very high noise. }
      \label{fig:ahvn}
  \end{figure}

    \begin{figure}
      \centering
      \includegraphics[width=1.\linewidth]{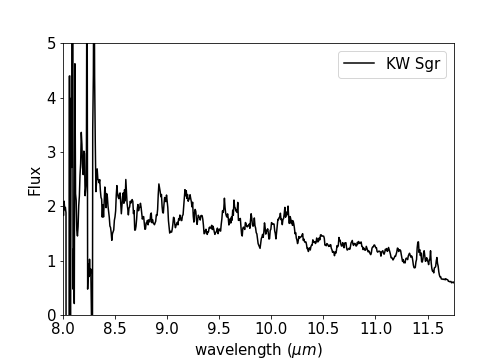}
      \caption{Same as Fig.~\ref{fig:ahfn} but for KW~Sgr.}
      \label{fig:fwfn}
  \end{figure}
  \begin{figure}
      \centering
      \includegraphics[width=1.\linewidth]{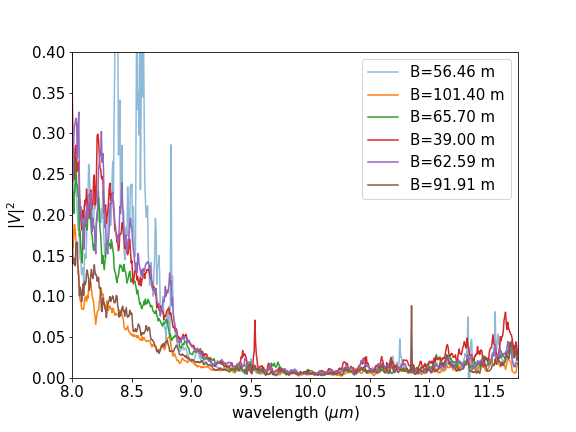}
      \caption{Same as Fig.~\ref{fig:ahvn} but for KW~Sgr.}
      \label{fig:kwvn}
  \end{figure}

      \begin{figure}
      \centering
      \includegraphics[width=1.\linewidth]{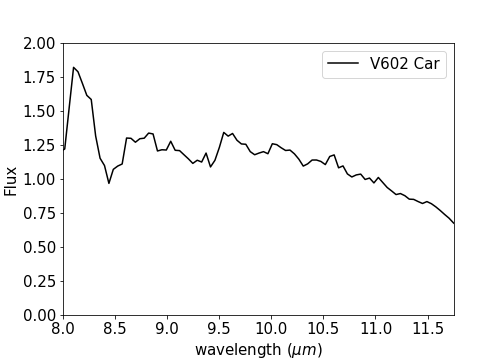}
      \caption{Same as Fig.~\ref{fig:ahfn} but for V602~Car. The resolution is low in this case.}
      \label{fig:v602fn}
  \end{figure}
  \begin{figure}
      \centering
      \includegraphics[width=1.\linewidth]{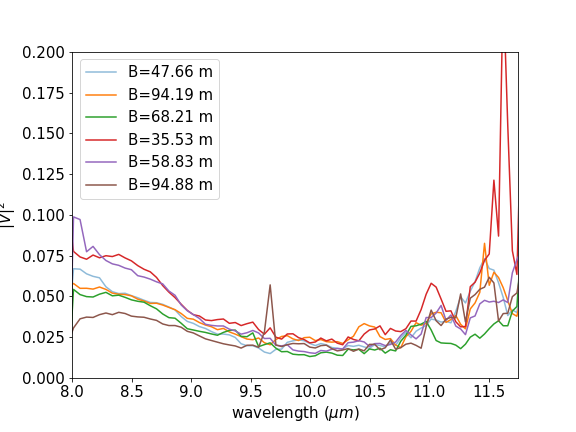}
      \caption{Same as Fig.~\ref{fig:ahvn} but for V602~Car. The resolution is low in this case.}
      \label{fig:v602vn}
  \end{figure}

        \begin{figure}
      \centering
      \includegraphics[width=1.\linewidth]{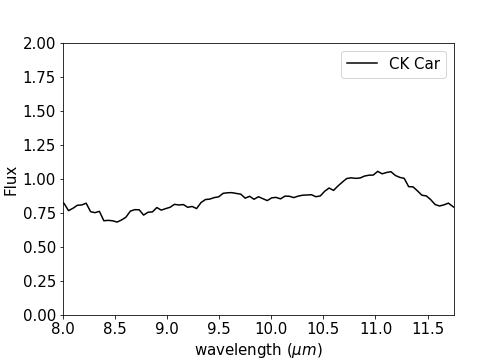}
      \caption{Same as Fig.~\ref{fig:ahfn} but for CK~Car. The resolution is low in this case.}
      \label{fig:vckfn}
  \end{figure}
  \begin{figure}
      \centering
      \includegraphics[width=1.\linewidth]{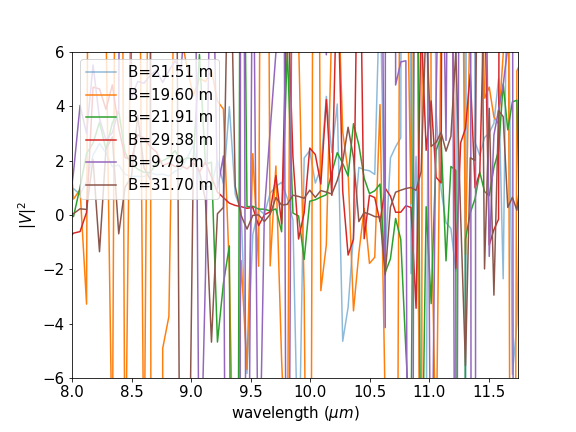}
      \caption{Same as Fig.~\ref{fig:ahvn} but for CK~Car. The resolution is low in this case.}
      \label{fig:vckvn}
  \end{figure}

          \begin{figure}
      \centering
      \includegraphics[width=1.\linewidth]{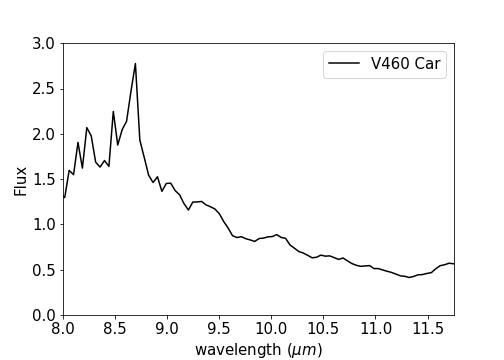}
      \caption{Same as Fig.~\ref{fig:ahfn} but for V460~Car. The resolution is low in this case.}
      \label{fig:v460fn}
  \end{figure}
  \begin{figure}
      \centering
      \includegraphics[width=1.\linewidth]{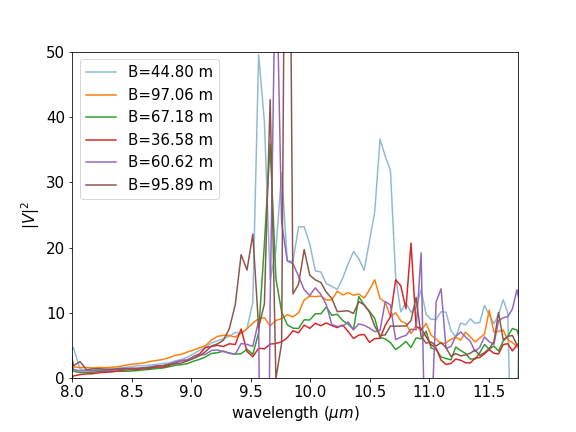}
      \caption{Same as Fig.~\ref{fig:ahvn} but for V460~Car. The resolution is low in this case.}
      \label{fig:v460vn}
  \end{figure}
\end{appendix}
\end{document}